\documentclass[prx,reprint,showpacs,superscriptaddress,longbibliography]{revtex4-1}
\usepackage{amsmath,amssymb,bm,mathrsfs,graphicx, braket,amsthm,enumerate,times}
\usepackage[colorlinks=true,citecolor=blue,linkcolor=blue]{hyperref}
\usepackage{longtable}
\usepackage{multirow}

\renewcommand{\vec}[1]{\bm{#1}}

\begin{document} 
\title{Inequivalent Berry phases for the bulk polarization}
\author{Haruki Watanabe} 
\email{haruki.watanabe@ap.t.u-tokyo.ac.jp}
\affiliation{Department of Applied Physics, University of Tokyo, Tokyo 113-8656, Japan}

\author{Masaki Oshikawa} 
\email{oshikawa@issp.u-tokyo.ac.jp}
\affiliation{Institute for Solid State Physics, University of Tokyo, Kashiwa 277-8581, Japan} 
\begin{abstract}
We discuss characterization of the polarization for insulators under the periodic boundary condition in terms of the Berry phase, clarifying confusing subtleties.  
For band insulators, the Berry phase can be formulated in terms of the Bloch function in the momentum space.  More generally, in the presence of interactions or disorders, one can instead use the many-body ground state as a function of the flux piercing the ring.  However, the definition of the Bloch function and the way describing the flux are not unique.  As a result, the value of the Berry phase and its behavior depend on how precisely it is defined. In particular, identifying the Berry phase as a polarization, its change represents a polarization current which also depends on the definition.
We demonstrate this by elucidating mutual relations among different definitions of the Berry phase, and that they correspond to the current measured differently in the real space.
Despite the non-uniqueness of the polarization current, 
the total charge transported during a Thouless pumping process is independent of the definition, reflecting its topological nature. 
\end{abstract}
\maketitle

\section{Introduction}
The polarization is of fundamental importance in understanding condensed-matter systems~\cite{AshcroftMermin,Kittel}.  
Historically, theories of the polarization were first developed in order to understand ferroelectric materials with macroscopic electric polarization~\cite{Rabe}. The total electric dipole moment of a piece of material may be given in terms of the surface charge. However, since the total electric dipole moment is typically proportional to the volume of the material, it could be regarded as a bulk property. Hence the electric polarization may be defined with some caveat for the system with the periodic boundary condition, for which the surface charge is absent.
It turned out that the concept of the polarization is useful in describing much wider materials and phenomena than the ferroelectricity.
For example, the spin transport in topological insulators can be understood via `spin polarization'~\cite{FuTRP,XiaoLiangQi}.
One of the key observations was the identification of the polarization as a Berry phase~\cite{Zak,KSVPRB1993,VanderbiltKingSmith,RestaRMP,RestaVanderbilt,OrtizMartin,AligiaOrtiz,Souza}, which revealed the topological nature of the polarization.
Topological transports such as the quantum Hall effect and the Thouless pump~\cite{Thouless,NiuThouless} are deeply related to the polarization since the Chern number can be understood in terms of the adiabatic evolution of the polarization~\cite{XiaoLiangQi}.
However, there is a substantial confusion in the very definition of the polarization as a Berry phase (see, e.g. Ref.~\onlinecite{Martin}). Several inequivalent Berry phases can be defined, and indeed found in the literature.

Given the fundamental importance of the polarization, in this paper, we revisit the relation between the polarization and the Berry phase. Our systematic analysis clarifies the physical meanings of different forms of the Berry phase.
As we will discuss in details, they can be related to the polarization, while only one particular definition of the Berry phase corresponds to the polarization which is standard in the literature. Nevertheless, other definitions of the Berry phase are also perfectly consistent and have their own physical meanings.
Although the ``polarization current'' derived from the Berry phase does depend on the definition, the total charge transported during a Thouless pumping is given by the same quantized topological invariant.

This paper is organized as follows.
In Sec.~\ref{sec:Thouless}, after reviewing the Thouless pump, we introduce two Berry phases, one for the uniform vector potential and the other for twisted boundary condition, and clarify their meaning and properties.
We confirm and demonstrate our understanding in a concrete model, in Sec.~\ref{sec:model}.
We then clarify the relation between the Berry-phase formulation of the polarization and the compact expression proposed by Resta in Sec.~\ref{sec:Resta}.
In Sec.~\ref{sec:band}, we discuss the special case of band insulators.
Finally, Sec.~\ref{sec:conclusion} is devoted to conclusions.

\section{General Formulation}
\label{sec:Thouless}
\subsection{Thouless pump}

In order to motivate the formulation,
let us start with reviewing the Thouless pump~\cite{Thouless,NiuThouless}. For simplicity, we discuss the quantum mechanics of particles on a 1D ring. In the Thouless pump, the Hamiltonian $\hat{H}(t)$ is adiabatically changed over time in such a way that $\hat{H}(0)=\hat{H}(T)$, and we consider the charge transported during the period $0 \leq t \leq T$.  Although the pumping itself can be realized just by the adiabatic time-dependence of the Hamiltonian, it is convenient for theoretical analysis to introduce a magnetic flux $\theta$ piercing the ring~\cite{NiuThouless}.
Let us represent the flux $\theta$ by the position and time-independent vector potential $A_x=\tfrac{\theta}{L}$.
Then the simplest example of the Hamiltonian reads
\begin{equation}
\hat{H}_\theta(t)=\int_{0}^L dx\,\hat{c}_x^\dagger\left[-\tfrac{1}{2m}(\partial_x+i\tfrac{\theta}{L})^2+V_x(t)\right]\hat{c}_x. \label{H1}
\end{equation}
Throughout this paper, we set the charge of the particle to unity.
Our discussion below does not rely on the specific form of the Hamiltonian~\eqref{H1}. Arbitrary finite-range interactions can be added as long as the particle number conservation is respected.

There is a tradeoff between the periodicity in $x$ and that in $\theta$.  The current choice of the uniform vector potential implicitly assumes the periodic boundary condition in space.  On the other hand, $\hat{H}_{\theta+2\pi}$ is not identical to $\hat{H}_\theta$ and is only unitarily equivalent to $\hat{H}_\theta$ although $\theta$ and $\theta+2\pi$ are physically equivalent.  (For the sake of brevity, we do not explicitly write the time dependence below when it is obvious.)  These two values of $\theta$'s are related by the large gauge transformation $e^{2\pi i\hat{P}}$ as $\hat{H}_{\theta+2\pi}=e^{-2\pi i\hat{P}}\hat{H}_{\theta}e^{2\pi i\hat{P}}$, where
\begin{equation}
\hat{P}\equiv\tfrac{1}{L}\int_{0}^L dx\,x\hat{n}_x,\quad \hat{n}_x\equiv\hat{c}_{x}^\dagger \hat{c}_{x}\label{Pdef}
\end{equation}
is the polarization operator.

Since the vector potential is uniform, taking a derivative of $\hat{H}_\theta$ with respect to $\theta$ gives the \emph{averaged} current operator,
\begin{eqnarray}
\hat{\bar{j}}_{\theta}\equiv \partial_\theta\hat{H}_\theta=\tfrac{1}{L}\int_{0}^Ldx\,\hat{j}_\theta(x).\label{Jave}
\end{eqnarray}
Here, $\hat{j}_\theta(x)$ is the local current. For instance, it reads $\hat{j}_\theta(x)=\tfrac{1}{2mi}\hat{c}_x^\dagger(\partial_x+i\tfrac{\theta}{L})\hat{c}_x+\text{h.c.}$ for the Hamiltonian in Eq.~\eqref{H1}.

Let us denote by $|\Phi_\theta\rangle$ the ground state of the snapshot Hamiltonian with the energy eigenvalue $E_\theta$.  We assume the uniqueness of the ground state $|\Phi_\theta\rangle$ and the finite excitation gap above the ground state for all values of $\theta\in[0,2\pi]$ and $t\in[0,T]$.  Let $|\Psi_\theta(t)\rangle$ be the state that is initially the ground state $|\Phi_\theta\rangle$ at $t=0$.  By taking into account the leading contribution of the excited states to $|\Psi_\theta(t)\rangle$ for $t>0$, Niu and Thouless showed that the current expectation value $\mathcal{J}_\theta(t)\equiv\langle\Psi_\theta(t)|\hat{\bar{j}}_{\theta}|\Psi_\theta(t)\rangle$ at each time $t$ is given in the form of the Berry curvature~\cite{NiuThouless}:
\begin{eqnarray}
\mathcal{J}_\theta&=&\partial_{\theta}E_\theta+\mathcal{F}_\theta,\label{NTcurrent}\\
\mathcal{F}_\theta&\equiv&i\big[\partial_t\langle\Phi_\theta|\partial_{\theta}|\Phi_\theta\rangle-\partial_{\theta}\langle\Phi_\theta|\partial_t|\Phi_\theta\rangle\big].\label{Ftheta}
\end{eqnarray}
We review the derivation in Appendix~\ref{app:thouless}.  The term $\partial_{\theta}E_\theta$ is the persistent current of the ground state that can be neglected for a large $L$.  Furthermore, Niu and Thouless also showed that $\mathcal{J}_\theta$ can be well-approximated by the average over $\theta$,
\begin{eqnarray}
\mathcal{J}&\equiv&\int_{0}^{2\pi}\tfrac{d\theta}{2\pi}\mathcal{J}_\theta=\int_{0}^{2\pi}\tfrac{d\theta}{2\pi}\mathcal{F}_\theta\notag\\
&=&\int_{0}^{2\pi}\tfrac{d\theta}{2\pi}i\big[\partial_t\langle\Phi_\theta|\partial_{\theta}|\Phi_\theta\rangle-\partial_{\theta}\langle\Phi_\theta|\partial_t|\Phi_\theta\rangle\big],\label{Jave2}
\end{eqnarray}
when $L$ is sufficiently large~\cite{NiuThouless}.  After all, the transported charge $Q\equiv\int_{0}^Tdt\mathcal{J}$ during this time period is given in the form of the Chern number~\cite{Thouless,NiuThouless}:
\begin{eqnarray}
C\equiv\int_{0}^Tdt\int_{0}^{2\pi}\tfrac{d\theta}{2\pi}\,\mathcal{F}_\theta,\label{tChern}
\end{eqnarray}
which reveals the topological nature of the pump and suggests the quantization of the transported charge (but see blow).

\subsection{Berry phase with uniform vector potential}

Physically, we demand that the polarization $\mathcal{P}$ satisfies
\begin{equation}
 \mathcal{J} = \tfrac{d}{dt} \mathcal{P} .
\label{JvsP}
\end{equation}
%This suggests the identification of the integral
Comparing Eq.~\eqref{JvsP} with the first term in the integrand of Eq.~\eqref{Jave2}, it is tempting to identify the integral
\begin{eqnarray}
\mathcal{P}\sim\int_{0}^{2\pi}\tfrac{d\theta}{2\pi}\,i\langle\Phi_\theta|\partial_{\theta}|\Phi_\theta\rangle\label{manybodyP}
\end{eqnarray}
as the polarization.
Indeed, Eq.~\eqref{manybodyP} is the standard definition of the polarization in the
bulk~\cite{OrtizMartin,AligiaOrtiz,Souza,Aligia,Hetenyi2012}, while there is a
subtlety as we will discuss below.
We note that, Eq.~\eqref{JvsP} and our convention of unit charge imply that $\mathcal{P}$ is dimensionless, which is consistent with Eq.~\eqref{manybodyP}.

The form~\eqref{manybodyP} looks like a Berry phase.
However, as we pointed out above, the Hamiltonian $\hat{H}_\theta$ lacks the periodicity in $\theta$, and thus the state $|\Phi_\theta\rangle$ is not periodic either. In fact, the value of Eq.~\eqref{manybodyP} can be arbitrarily modified by the gauge transformation $|\Phi_\theta\rangle\rightarrow |\Phi_\theta\rangle'=e^{i\chi(\theta)}|\Phi_\theta\rangle$ that would shift $\mathcal{P}$ by $\frac{\chi(2\pi)-\chi(0)}{2\pi}$.
We thus need to define the polarization as
\begin{eqnarray}
\mathcal{P}\equiv\int_{0}^{2\pi}\tfrac{d\theta}{2\pi}\,i\langle\Phi_\theta|\partial_{\theta}|\Phi_\theta\rangle+\frac{1}{2\pi}\text{Im}\ln\langle\Phi_{0}|e^{2\pi i \hat{P}}|\Phi_{2\pi}\rangle
\label{manybodyP12},
\end{eqnarray}
whose fractional part can be confirmed as gauge invariant.
In fact, one can reproduce both the first and the second term in the integrand of Eq.~\eqref{Jave2} by plugging Eq.~\eqref{manybodyP12} into Eq.~\eqref{JvsP} using $|\Phi_{2\pi}\rangle=e^{i\alpha(t)}e^{-2\pi i \hat{P}}|\Phi_{0}\rangle$ for some $\alpha(t)\in[0,2\pi]$.
However, the topological nature of the polarization is not obvious
in this formulation.
In fact, even though Eq.~\eqref{tChern} appears as a Chern number,
the lack of the periodicity in $\theta$ would invalidate the usual
argument of its quantization.
Thus we need to study the issue more carefully.

\subsection{Berry phase under twisted boundary condition}
\label{sec:twisted}

To make the topological quantization evident, let us perform the unitary transformation
\begin{equation}
\hat{\tilde{H}}_{\theta}=e^{ i\theta\hat{P}}\hat{H}_{\theta}e^{- i\theta \hat{P}}.
\end{equation}
The new Hamiltonian has the nice periodicity in $\theta$, $\hat{\tilde{H}}_{\theta+2\pi}=\hat{\tilde{H}}_{\theta}$, but instead the boundary condition is twisted by the factor $e^{i\theta}$ (See Appendix~\ref{app:twisted}).  Let $|\tilde{\Phi}_\theta\rangle$ be the unique ground state of $\hat{\tilde{H}}_{\theta}$.  As the Hamiltonian is periodic in $\theta$, one can naturally demand $|\tilde{\Phi}_{\theta+2\pi}\rangle=|\tilde{\Phi}_\theta\rangle$ without loss of the generality. Given $|\tilde{\Phi}_\theta\rangle$ with this property, we can fix the phase ambiguity of $|\Phi_{\theta}\rangle$ in the uniform gauge by setting
\begin{equation}
|\Phi_{\theta}\rangle=e^{-i\theta \hat{P}}|\tilde{\Phi}_\theta\rangle.
\label{rule}
\end{equation}
With this condition, $|\Phi_{\theta+2\pi}\rangle$ is related to $|\Phi_\theta\rangle$ as $|\Phi_{\theta+2\pi}\rangle=e^{-2\pi i \hat{P}}|\Phi_\theta\rangle$.  (In other words, $\alpha(t)$ above is set $0$.)
Then the second term in Eq.~\eqref{manybodyP12} vanishes and the definition of $\mathcal{P}$ reduces back to Eq.~\eqref{manybodyP}.
Furthermore, the gauge transformation consistent with this condition must satisfy $e^{i\chi(0)}=e^{i\chi(2\pi)}$ and the fractional part of $\mathcal{P}$ is gauge invariant.  The same condition also demands that $\langle \Phi_{2\pi} | \partial_t | \Phi_{2\pi} \rangle = \langle \Phi_{0} | \partial_t | \Phi_{0} \rangle$ so that $\int_0^{2\pi} d\theta\partial_\theta[\langle \Phi_{\theta} | \partial_t | \Phi_{\theta} \rangle]$ vanishes and $\mathcal{J}=\frac{d}{dt}\mathcal{P}$ precisely holds.

On the other hand, using $|\tilde{\Phi}_\theta\rangle$ instead of $|\Phi_\theta\rangle$, one may introduce a different kind of Berry phase~\cite{HiranoKatsuraHatsugai1,HiranoKatsuraHatsugai2}
\begin{eqnarray}
\tilde{\mathcal{P}}=\int_{0}^{2\pi}\tfrac{d\theta}{2\pi}\,i\langle\tilde{\Phi}_\theta|\partial_{\theta}|\tilde{\Phi}_\theta\rangle.
\label{tildeP}
\end{eqnarray}
It is tempting to identify this Berry phase as the polarization.
However, we find that even \emph{the fractional part} of $\mathcal{P}$ and $\tilde{\mathcal{P}}$ do not agree in general. Instead, Eq.~\eqref{rule} suggests that
\begin{equation}
\mathcal{P}=\tilde{\mathcal{P}}+\bar{\mathcal{P}}_0,\quad\bar{\mathcal{P}}_0\equiv\int_{0}^{2\pi}\tfrac{d\theta}{2\pi}\,\langle \Phi_{\theta}|\hat{P}|\Phi_{\theta}\rangle.\label{relP}
\end{equation}
The definition of $\bar{\mathcal{P}}_0$ here involves averaging over $\theta$, but it is exponentially close to the one without the average $\mathcal{P}_0\equiv\langle \Phi|\hat{P}|\Phi\rangle$ for a sufficiently large $L$~\cite{flux}.  Unlike $\mathcal{P}$ or $\tilde{\mathcal{P}}$, $\bar{\mathcal{P}}_0$ is completely gauge-independent. Ref.~\onlinecite{HatsugaiFukui} argued that $\mathcal{P}_0$ is related to the center of the mass position when the open boundary condition is taken.

To understand the physical meaning of the polarization-like quantity $\tilde{\mathcal{P}}$, note that
\begin{equation}
\partial_\theta\hat{\tilde{H}}_{\theta}=\hat{\tilde{j}}_\theta(0)\label{current2}
\end{equation}
is the \emph{local} current operator at the `seam' $x=0$ ($=L$).  This can be best seen by the fact that the unitary transformation $e^{i\theta\hat{P}}$ induces the gauge transformation $\tilde{A}_x=A_x-\partial_x(\frac{\theta}{L}x)=\tfrac{\theta}{L}-\frac{\theta}{L}(1-L\theta\delta(x))=\theta\delta(x)$. The delta function originates from the jump of $x$ by $-L$ at the seam.  As a sanity check, we have $\int_0^LdxA_x=\int_0^Ldx\tilde{A}_x=\theta$, which is required since the total flux piercing the ring should not be altered by the unitary transformation.  

Another way of verifying Eq.~\eqref{current2} is based on the current conservation law: $i[\hat{H}_\theta,\hat{n}_x]+\partial_x\hat{j}_\theta(x)=0$ (Appendix~\ref{app:conservation}). Plugging the definition of $\hat{P}$ in Eq.~\eqref{Pdef} and integrating by part, we get
\begin{eqnarray}
i[\hat{H}_\theta,\hat{P}]=-\tfrac{1}{L}\int_{0}^L dx\,x\partial_x\hat{j}_\theta(x)=\hat{\bar{j}}_{\theta}-\hat{j}_\theta(0),\label{dtP}
\end{eqnarray}
where $\hat{\bar{j}}_{\theta}$ is defined in Eq.~\eqref{Jave}. Therefore,
\begin{eqnarray}
\partial_\theta\hat{\tilde{H}}_{\theta}&=&\partial_\theta(e^{ i\theta\hat{P}}\hat{H}_{\theta}e^{- i\theta \hat{P}})\notag\\
&=&e^{ i\theta\hat{P}}\left(\partial_\theta\hat{H}_{\theta}-i[\hat{H}_\theta,\hat{P}]\right)e^{- i\theta \hat{P}}\notag\\
&=&e^{ i\theta\hat{P}}\hat{j}_\theta(0)e^{- i\theta \hat{P}}=\hat{\tilde{j}}_\theta(0).
\end{eqnarray}
The relation in Eq.~\eqref{dtP} is somewhat nontrivial --- in the Heisenberg picture, the left-hand side is $\partial_t\hat{P}$.  It means that $\hat{\bar{j}}_{\theta}\neq \partial_t\hat{P}$ at the operator level under the periodic boundary condition, although we still have $\mathcal{J}=\frac{d}{dt}\mathcal{P}$.

Given Eq.~\eqref{current2}, following the discussion of Niu-Thouless, we find that the expectation value $\tilde{\mathcal{J}}_\theta\equiv\langle\Psi_\theta(t)|\hat{\tilde{j}}_\theta(x=0)|\Psi_\theta(t)\rangle$ of the local current flowing at the seam, induced by the adiabatic time evolution, is given by
\begin{eqnarray}
\mathcal{\tilde{J}}_\theta&=&\partial_{\theta}E_\theta+\mathcal{\tilde{F}}_\theta,\label{NTcurrent2}\\
\mathcal{\tilde{F}}_\theta&\equiv&i\big[\partial_t\langle\tilde{\Phi}_\theta|\partial_{\theta}|\tilde{\Phi}_\theta\rangle-\partial_{\theta}\langle\tilde{\Phi}_\theta|\partial_t|\tilde{\Phi}_\theta\rangle\big],
\end{eqnarray}
It is now clear that \emph{$\tilde{\mathcal{P}}$ counts the number of particles that go through the seam.}  This is in sharp contrast to $\mathcal{P}$ that cares the motion of particles at every single point of space on the same footing, as suggested by Eq.~\eqref{Jave}.

\subsection{``Gauge'' dependence}
We have clarified that the polarization currents $\mathcal{J}$ and $\mathcal{\tilde{J}}$, respectively corresponding to $A_x=\frac{\theta}{L}$ and $\tilde{A}_x=\theta\delta(x)$, represent quite different quantities. In Sec.~\ref{sec:model}, we will demonstrate the clear difference between them in a simple example.  However, this might sound puzzling because they were written in the form of the Berry curvatures:
\begin{eqnarray}
\mathcal{J}=\tfrac{d}{dt}\mathcal{P}=\int_{0}^{2\pi}\tfrac{d\theta}{2\pi}\mathcal{F}_\theta,\\
\mathcal{\tilde{J}}=\tfrac{d}{dt}\mathcal{\tilde{P}}=\int_{0}^{2\pi}\tfrac{d\theta}{2\pi}\tilde{\mathcal{F}}_\theta,
\end{eqnarray}
and the Berry curvature must be ``gauge-invariant''.

\begin{figure}
\begin{center}
\includegraphics[width=0.99\columnwidth]{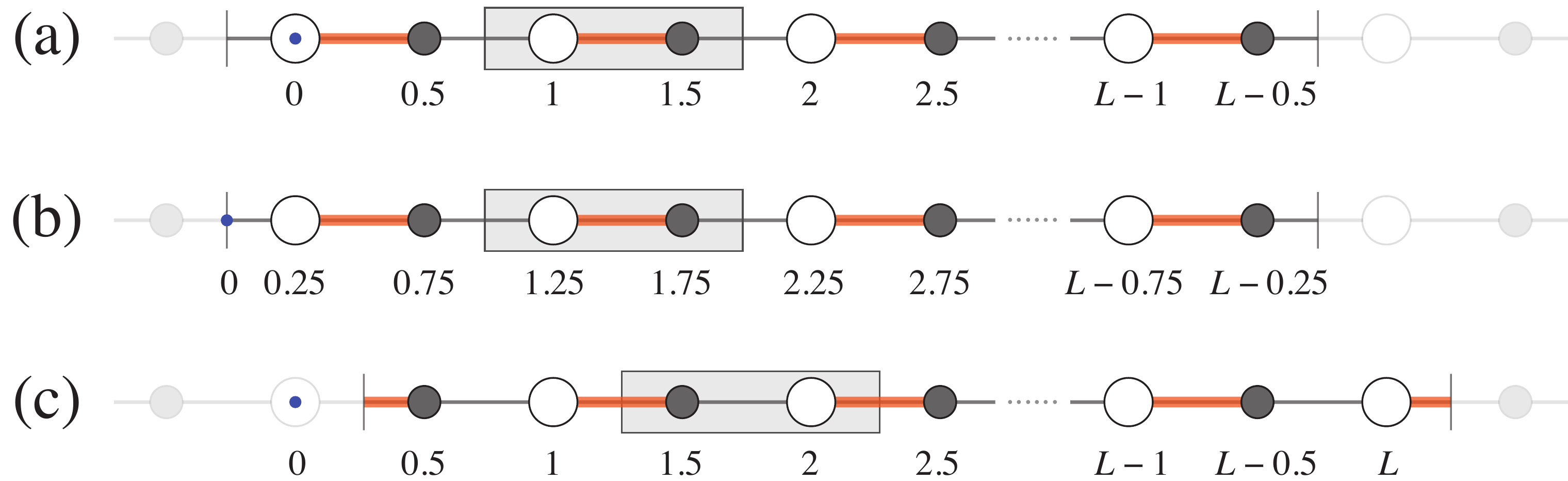}
\caption{\label{fig:1Dlattice} (a) A lattice with two sites ($x_1=0$ and $x_2=0.5$) in a unit cell. The lattice constant $a$ is set $1$. (b) The origin (shown by blue dot) is shifted by $-0.25$ and the lattice position becomes $x_1=0.25$ and $x_2=0.75$.  (c) A different unit cell is chosen, which includes $x_1=1$ and $x_2=0.5$. }
\end{center}
\end{figure}

To resolve this apparent paradox, one should note that there are two completely distinct types of gauge choices here. One is the choice of the vector potential $A_x$ associated with the local U(1) phase of the wavefunction \emph{as a function of $x$}. The gauge transformation in this sense is represented by the unitary operator $\hat{U}_{\epsilon}=e^{i\int dx\,\epsilon(x)\hat{n}_x}$ that induces $A_x\rightarrow A_x-\partial_x\epsilon(x)$ (Appendix~\ref{app:conservation}). The other one is the choice of the overall phase of the state vector. Since $|\Phi_\theta\rangle$ is defined for the snapshot Hamiltonian $\hat{H}_\theta$ independently for each $\theta$, we can always redefine $|\Phi_\theta\rangle'=e^{i\chi(\theta)}|\Phi_\theta\rangle$ \emph{as a function of $\theta$}.  The Berry curvature $\mathcal{F}_\theta$, $\tilde{\mathcal{F}}_\theta$ are independent of such a gauge choice in the $\theta$-space~\cite{Kohmoto} but may change under an $x$-dependent gauge transformation discussed above.  In fact, Eq.~\eqref{rule} implies
\begin{eqnarray}
\mathcal{F}_\theta-\tilde{\mathcal{F}}_\theta=\partial_t[\langle \Phi_{\theta}|\hat{P}|\Phi_{\theta}\rangle],\label{FF}
\end{eqnarray}
which is generically non-vanishing.

Although we have only compared the two representative choices of the vector potential so far,  one can freely move the seam or even split it (i.e., $A_x(x)=\theta\sum_{i}p_i\delta(x-x_i)$ with $\sum_ip_i=1$) by a proper local gauge transformation.  The corresponding Berry phase simply denotes the weighted average of the number of particles going through each seam.

\subsection{Issues in the polarization operator}
\label{sec:issues}

Although $\hat{P}$ in Eq.~\eqref{Pdef} is perfectly well-defined as it is, it has two unfavorable properties. (i) Origin dependence~\cite{Moore}: when the origin is shifted by $-\xi$ and $x$ is replaced with $x+\xi$, $\hat{P}$ becomes $\hat{P}'=\hat{P}+\xi\frac{\hat{N}}{L}$ [see Fig.~\ref{fig:1Dlattice}(b)].  (ii) Seam dependence: if the position of the seam is moved by $r$ and we use $r\leq x<L+r$ as the range of $x$, instead of $0\leq x<L$, $\hat{P}$ becomes $\hat{P}'=\hat{P}+\int_0^r dx\,\hat{n}_x$ [see Fig.~\ref{fig:1Dlattice}(c)].  As a consequence, $\bar{\mathcal{P}}_0$ depends both on the choice of the origin and the position of the seam. As we will see later, the choice of the position of the seam corresponds to the choice of the unit-cell~\cite{VanderbiltKingSmith,RestaVanderbilt} in the case of band insulators.

The origin dependence may be resolved by imposing the charge neutrality condition and taking into account contributions from all `charged' particles (e.g., ions for the charge polarization)~\cite{RestaVanderbilt}. However, if we understand the polarization in a generalized sense, including the spin polarization for $S_z$ conserving magnets~\cite{HiranoKatsuraHatsugai1}, the neutrality condition is not necessarily satisfied.

Since $\tilde{\mathcal{P}}$ cares only the position of the seam, it is independent of the choice of origin.  This implies that $\mathcal{P}=\tilde{\mathcal{P}}+\bar{\mathcal{P}}_0$ depends on the origin but is independent of the seam.   We summarize these properties in the Table~\ref{summary1}.

Another related but distinct issue in $\hat{P}$ is about the boundary condition~\cite{RestaPRL1998}.  When $|\Phi\rangle$ satisfies the periodic boundary condition,  $\hat{P}|\Phi\rangle$ does not because $\hat{P}$ multiplies $x$ to the wavefunction, which becomes discontinuous at the seam.  For this reason, strictly speaking, the quantity $\bar{\mathcal{P}}_0$ may not be an expectation value of an operator in the usual sense --- it appeared above as the difference of the two Berry phases with respect to the state under different boundary conditions. Nonetheless, since $\mathcal{P}$ and $\tilde{\mathcal{P}}$ are well-defined, $\bar{\mathcal{P}}_0=\mathcal{P}-\tilde{\mathcal{P}}$ should also be.

\begin{table}
\caption{Properties of the many-body Berry phases $\mathcal{P}$ and $\tilde{\mathcal{P}}$ defined in Eqs.~\eqref{manybodyP12} and \eqref{tildeP}, $\bar{\mathcal{P}}_{0}$ in Eq.~\eqref{relP}, and the change of Berry phases in Eqs.~\eqref{dp1} and \eqref{dp2}.
\label{summary1}}
\begin{tabular}{c|ccc}\hline\hline
$\mathcal{P}$ & Gauge\footnote{The `gauge' here refers to the gauge choice in the $\theta$-space.} & Origin& Seam\\\hline
$\mathcal{P}$ & depends\footnote{The fractional part of $\mathcal{P}$ and $\tilde{\mathcal{P}}$ is gauge-independent.}  &depends\footnote{The origin dependence may be resolved by the charge-neutrality condition.}& independent \\
$\tilde{\mathcal{P}}$ & depends$^{\text{\textcolor{blue}{b}}}$ &  independent   & depends \\
$\bar{\mathcal{P}}_{0}$ & independent & depends$^{\text{\textcolor{blue}{c}}}$ & depends \\\hline
$\Delta\mathcal{P}(t)$ & independent & independent & independent \\
$\Delta\tilde{\mathcal{P}}(t)$ & independent  &  independent   & depends \\\hline\hline
\end{tabular}
\end{table}

\subsection{Change of the polarization}
In order to cancel out the dependence on the unphysical quantities, it is customary to focus on the difference,
\begin{eqnarray}
\Delta\mathcal{P}(t)&\equiv&\mathcal{P}(t)-\mathcal{P}(0)=\int_{0}^{t}dt'\int_{0}^{2\pi}\tfrac{d\theta}{2\pi}\mathcal{F}_\theta(t'),\label{dp1}\\
\Delta\tilde{\mathcal{P}}(t)&\equiv&\tilde{\mathcal{P}}(t)-\tilde{\mathcal{P}}(0)=\int_{0}^{t}dt'\int_{0}^{2\pi}\tfrac{d\theta}{2\pi}\tilde{\mathcal{F}}_\theta(t').\label{dp2}
\end{eqnarray}
Neither $\Delta\mathcal{P}(t)$ or $\Delta\tilde{\mathcal{P}}(t)$ depends on the choice of the origin. Note, however, that $\Delta\tilde{\mathcal{P}}$ still depends on the position of the seam because, by definition, it measures the current flowing at the seam.

Despite $\Delta\tilde{\mathcal{P}}(t)\neq \Delta\mathcal{P}(t)$ in general, when $t$ is the period $T$ of the cyclic evolution of the Hamiltonian, we have
\begin{equation}
\Delta\tilde{\mathcal{P}}(T)=\Delta\mathcal{P}(T)=Q.
\end{equation}
Namely, the quantized charge transport is independent of the choice of the vector potential $A_x$.  This can be readily seen based on Eq.~\eqref{FF}.  Although $|\Phi_{\theta}\rangle$ at $t=0$ and $T$ may differ by a phase, the expectation value $\langle \Phi_{\theta}|\hat{P}|\Phi_{\theta}\rangle$ should be manifestly periodic in $t$ and the total derivation does not contribute to the integral $\int_{0}^{T}dt$.

\subsection{Higher dimensions}

Before moving on to our the analysis of a concrete model, let us comment on how to generalize our formulae to higher dimensions.   As Berry phases are essentially one-dimensional quantity, we have formulated them in 1D systems. To extend them to higher-dimensions, we should take periodic boundary conditions in all directions with the period $L_i$ ($i=1,\cdots,d$). Correspondingly, the integral $\int dx$ in Eqs.~\eqref{Pdef} and \eqref{Jave} should be replaced by $\int d^d\vec{x}$:
\begin{eqnarray}
\hat{P}&\equiv&\tfrac{1}{L_1}\int d^d\vec{x}\,x\hat{n}_{\vec{x}},\\
\hat{\bar{j}}_{\theta}&\equiv&\partial_\theta\hat{H}_\theta=\tfrac{1}{L_1}\int d^d\vec{x}\,\hat{j}_\theta(\vec{x}).
\end{eqnarray}
In order to identify $e^{2\pi i \hat{P}}$ as the large gauge transformation operator, we \emph{do not} replace the $L_1^{-1}$ factor by $(L_1L_2\cdots L_d)^{-1}$.

\begin{figure}
\begin{center}
\includegraphics[width=0.80\columnwidth]{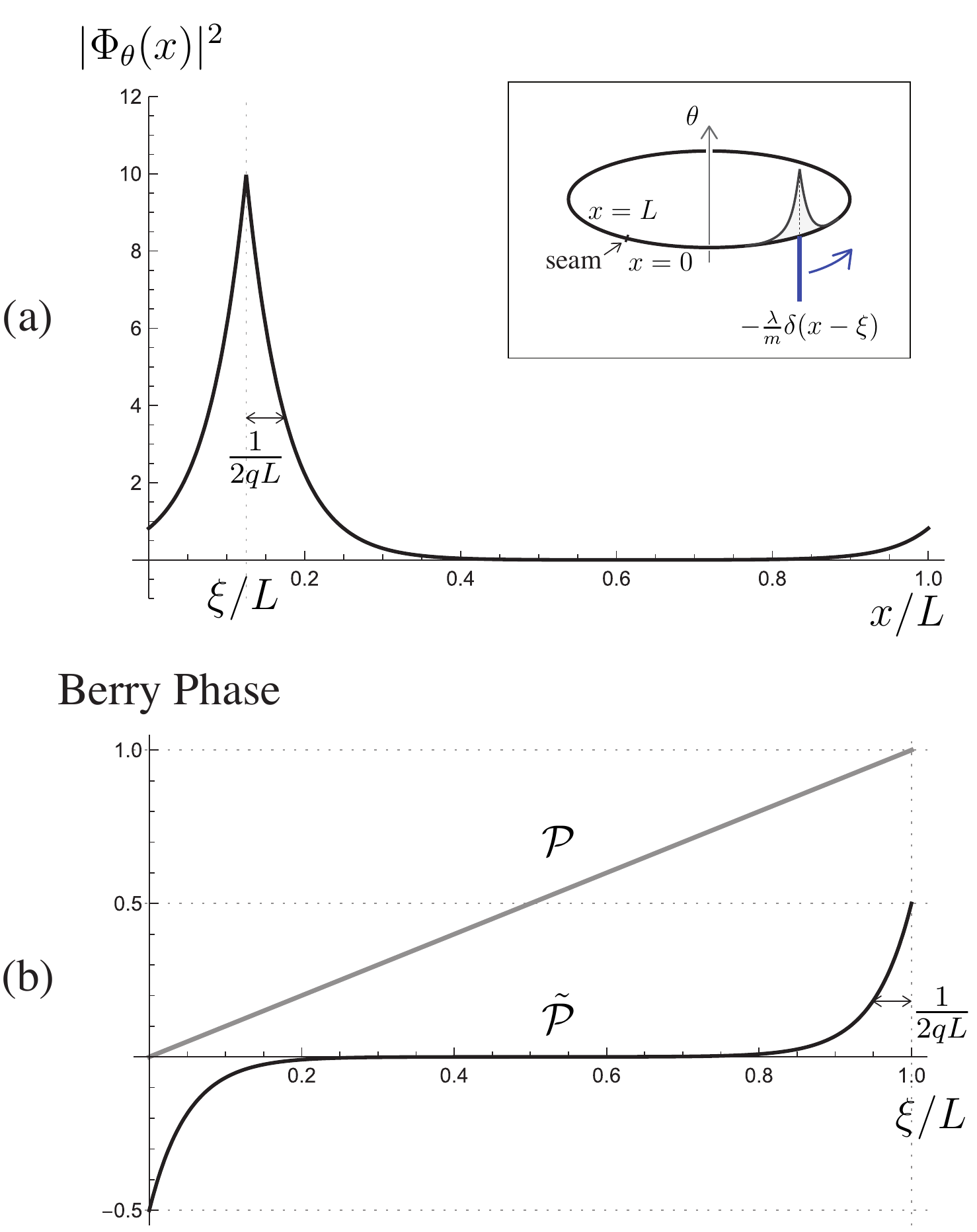}
\caption{\label{fig:Delta} (a) The probability amplitude for $q L=10$, $\xi=\frac{1}{8}L$, and $\theta=0$. The inset illustrates the setup.  (b) $\mathcal{P}$ (gray) and $\tilde{\mathcal{P}}$ (black) as a function of $\xi$ for $q L=10$.}
\end{center}
\end{figure}

\section{Model with delta-function potential.}
\label{sec:model}

Let us confirm this understanding through a simple \emph{one-particle} model in one dimension. We take the delta-function potential $V_x=-\frac{\lambda}{m}\delta(x-\xi)$ centered at $x=\xi$ in Eq.~\eqref{H1}.  The Hamiltonian has a unique bound state with the negative energy $E_\theta=-\frac{q^2}{2m}$, where $q\simeq \lambda$ should be found by inverting $\lambda=q\tfrac{\cosh q L-\cos\theta}{\sinh q L}$.  Under the periodic boundary condition $\Phi_\theta(L)=\Phi_\theta(0)$, the ground-state wavefunction, satisfying $\Phi_{\theta+2\pi}(x)=e^{-i\frac{2\pi}{L}x}\Phi_\theta(x)$, is given by
\begin{eqnarray}
\Phi_\theta(x)&=&\mathcal{N}_\theta e^{-i\theta\frac{x}{L}}\Big[e^{-q|x-\xi|}(1-e^{-q L+i\theta\text{sgn}(x-\xi)})\notag\\
&&+e^{+q|x-\xi|}(e^{-q L+i\theta\text{sgn}(x-\xi)}-e^{-2q L})\Big],
\end{eqnarray}
where $\mathcal{N}_\theta$ is the normalization factor.  Other eigenenergies are all positive so that the excitation gap remains finite for any $\theta$. 

Now, suppose $\xi$ has a weak time-dependence, adiabatically increasing from $\xi=0$ at $t=0$ to $\xi=L$ at $t=T$.  We find
\begin{equation}
\mathcal{P}=i\int_{0}^{2\pi}\tfrac{d\theta}{2\pi}\int_0^Ldx\,\Phi_\theta(x)^*\partial_\theta\Phi_\theta(x)=\tfrac{\xi}{L}.
\end{equation}
The transported charge is thus $Q=\Delta\mathcal{P}(T)=+1$.  See the grey straight line in Fig.~\ref{fig:Delta}(b).

After the gauge transformation, the wavefunction becomes $\tilde{\Phi}_\theta(x)=e^{i\theta\frac{x}{L}}\Phi_\theta(x)$. It satisfies the twisted boundary condition $\tilde{\Phi}_\theta(L)=e^{i\theta}\tilde{\Phi}_\theta(0)$ and is periodic in $\theta$, $\tilde{\Phi}_{\theta+2\pi}(x)=\tilde{\Phi}_\theta(x)$. When $q L\gg1$, the Berry phase $\tilde{\mathcal{P}}$ is well-approximated by
\begin{eqnarray}
\tilde{\mathcal{P}}&=&i\int_{0}^{2\pi}\tfrac{d\theta}{2\pi}\int_0^Ldx\,\tilde{\Phi}_\theta(x)^*\partial_\theta\tilde{\Phi}_\theta(x)\notag\\
&\simeq&\tfrac{1}{2}(e^{2q (\xi-L)}-e^{-2q \xi}).
\end{eqnarray}
See the black curve in Fig.~\ref{fig:Delta}(b).  The exact expression is included in the Appendix~\ref{app:delta}.  As $q L$ increases, the slope of $\tilde{\mathcal{P}}$ near $\xi=0$ and $L$ becomes sharper and sharper. In the limit of $q L\rightarrow\infty$ (i.e., the tight-binding limit), $\tilde{\mathcal{P}}$ vanishes for $0<\xi<L$ and the jump at $\xi=0$ and $L$ becomes abrupt just like the step-function. 
Regardless of the values of $q L$, we see that $Q=\Delta\tilde{\mathcal{P}}(T)=+1$.   This behavior is perfectly consistent with our interpretation of $\tilde{\mathcal{P}}$ explained above --- only the motion across the seam affects $\tilde{\mathcal{P}}$.  We plot $\mathcal{P}$ as a function of $\xi$ for several other choices of the vector potential in Appendix~\ref{app:delta}.

An alternative way of viewing this particular model is via the Aharonov-Bohm phase. The wavefunction $\tilde{\tilde{\Phi}}_\theta(x)\equiv e^{i\theta\frac{\xi}{L}}\Phi_\theta(x)$ possesses the periodicity in $\xi$. For fixed $\theta$, the Berry phase $\mathcal{B}(\theta)\equiv-i\int_0^Ld\xi\int_0^Ldx\,\tilde{\tilde{\Phi}}_\theta(x)^*\partial_\xi\tilde{\tilde{\Phi}}_\theta(x)$ with respect to $\xi$ measures the flux piercing the ring $\theta$~\cite{Berry} in the limit of $q L\gg1$ and the difference $\frac{\mathcal{B}(2\pi)-\mathcal{B}(0)}{2\pi}=+1$ counts the transported charge. In this picture, the gauge-independence of transported charge is manifest because the Aharonov-Bohm phase is gauge-independent.

\section{Resta's formula}
\label{sec:Resta}

Let us clarify the relation between the Berry phases introduced above and the compact expression for the polarization proposed by Resta. It is formulated in terms of the expectation value of the large gauge transformation $e^{2\pi i\hat{P}}$ on the ground state $|\Phi\rangle\equiv|\Phi_{\theta=0}\rangle=|\tilde{\Phi}_{\theta=0}\rangle$~\cite{RestaPRL1998}:
\begin{equation}
\mathcal{P}_{\text{R}}\equiv\tfrac{1}{2\pi}\text{Im}\ln \langle\Phi|e^{2\pi i\hat{P}}|\Phi\rangle,\label{PR}
\end{equation}
Just like $\mathcal{P}$ and $\tilde{\mathcal{P}}$, only the fractional part of $\mathcal{P}_{\text{R}}$ is well-defined because of the ambiguity in the logarithm.  
While the formula~\eqref{PR} was originally introduced for one-dimensional systems, it might appear that the same formula can be straightforwardly used in any dimension. However, a care must be taken in higher dimensions (as it was hinted in Ref.~\onlinecite{RestaPRL1998}). We will clarify the issue below. 

Under the condition in Eq.~\eqref{rule}, the expectation value $\langle\Phi|e^{2\pi i\hat{P}}|\Phi\rangle$ can be interpreted as the overlap of $|\Phi_{\theta}\rangle$ at $\theta=0$ and $2\pi$~\cite{NakamuraVoit}.  We convert it into the form of the Berry phase by interpolating $\theta_n=\frac{2\pi n}{N}$ ($n=1,\cdots,N-1$):
\begin{eqnarray}
\langle\Phi_{2\pi}|\Phi_{0}\rangle&\simeq&\langle\Phi_{2\pi}|\Phi_{\pi}\rangle\langle\Phi_{\pi}|\Phi_{0}\rangle\notag\\
&\simeq&\langle\Phi_{2\pi}|\Phi_{\frac{3}{2}\pi}\rangle\langle\Phi_{\frac{3}{2}\pi}|\Phi_{\pi}\rangle\langle\Phi_{\pi}|\Phi_{\frac{1}{2}\pi}\rangle\langle\Phi_{\frac{1}{2}\pi}|\Phi_{0}\rangle\notag\\
&\simeq&\prod_{n=0}^{N-1}\langle \Phi_{\theta_{n+1}}|\Phi_{\theta_n}\rangle,\quad N=2^M.
\end{eqnarray}
In Appendix~\ref{app:Resta}, we prove that 
\begin{eqnarray}
\Big|\langle\Phi_{2\pi}|\Phi_{0}\rangle-\lim_{M\rightarrow\infty}\prod_{n=0}^{2^M-1}\langle \Phi_{\theta_{n+1}}|\Phi_{\theta_n}\rangle\Big|\leq\tfrac{(2\pi)^2\mathcal{C}V}{2\Delta^2L_1^2},
\label{PRbound}
\end{eqnarray}
where $V=L_1L_2\cdots L_d$ is the volume of the system, $\Delta=\min_{\theta}\Delta_\theta$ and $\Delta_\theta$ is the excitation gap of $\hat{H}_\theta$, and $\mathcal{C}$ is the current fluctuation defined by
\begin{equation}
\mathcal{C}\equiv \tfrac{L_1^2}{V}\max_\theta\langle\Phi_{\theta}|(\delta\hat{\bar{j}}_\theta)^2|\Phi_{\theta}\rangle,\,\,\,\delta\hat{\bar{j}}_\theta\equiv\hat{\bar{j}}_\theta-\langle\Phi_{\theta}|\hat{\bar{j}}_\theta|\Phi_{\theta}\rangle.
\end{equation}
In gapped phases where correlation functions decay exponentially, $\mathcal{C}$ converges to a finite $O(1)$ number in the limit of large system size. After the interpolation, the overlap is precisely the Berry phase:
\begin{eqnarray}
&&\lim_{N\rightarrow\infty}\prod_{n=0}^{N-1}\langle \Phi_{\theta_{n+1}}|\Phi_{\theta_n}\rangle\notag\\
&&=\lim_{N\rightarrow\infty}e^{-\sum_{n=0}^{N-1}\frac{2\pi}{N}\langle \Phi_{\theta}|\partial_\theta|\Phi_{\theta}\rangle|_{\theta=\theta_n}}=e^{2\pi i\mathcal{P}}.
\end{eqnarray}
Therefore, we have
\begin{equation}
\mathcal{P}_{\text{R}}=\tfrac{1}{2\pi}\text{Im}\ln\langle\Phi_{2\pi}|\Phi_{0}\rangle=\mathcal{P}
\end{equation}
if $\frac{V}{L_1^2}=\frac{L_2\cdots L_d}{L_1}\rightarrow0$.  
This condition is violated in dimensions $d\geq2$ when the thermodynamic limit is taken in the isotropic manner ($L_i = L$), but can be satisfied in some anisotropic cases.  For example, Ref.~\cite{Nakagawa} considered $\mathcal{P}_{\text{R}}$ in the thin-torus limit of 2D system and there $\frac{V}{L_1^2}=\frac{L_2}{L_1}\rightarrow0$ holds.

Resta's original argument~\cite{RestaPRL1998} relating $\mathcal{P}_{\text{R}}$ to the polarization is via an adiabatic time evolution. 
He introduced a weak time dependence and showed that $\frac{d}{dt}\mathcal{P}_{\text{R}}$ coincides with $\mathcal{J}_{\theta=0}$, defined in Eq.~\eqref{NTcurrent}, to the leading order in $L_1^{-1}$. However, his argument is based on the first-order perturbation theory, expanding $|\Phi_{\theta+d\theta}\rangle$ as $|\Phi_{\theta}\rangle+d\theta |\Phi_{\theta}^{(1)}\rangle+\cdots$ for ``$d\theta=2\pi$''. Such an expansion cannot be verified in general.  
In appendix~\ref{app:Resta}, we find that $1-|\langle\Phi_{\theta}|\Phi_{\theta+d\theta}\rangle|^2\leq (d\theta)^2\frac{\mathcal{C}V}{\Delta^2L_1^2}$ to the leading order in $d\theta$; when the right-hand side is small, $|\Phi_{\theta+d\theta}\rangle$ should be close to $|\Phi_{\theta}\rangle$ and the perturbation may be well-controlled. This condition is violated when $d\theta=2\pi$ and $\frac{\mathcal{C}V}{\Delta^2L_1^2}>1$.

While Eq.~\eqref{PRbound} is an inequality, we generally expect that
\begin{eqnarray}
\Big|\langle\Phi_{2\pi}|\Phi_{0}\rangle -  e^{i 2 \pi \mathcal{P}} \Big|
  \propto \tfrac{V}{\Delta^2 L_1^2}.
\end{eqnarray}
In fact, one-dimensional insulators, Resta and Sorella showed that~\cite{RestaPRL1999}
\begin{equation}
|\langle\Phi^{\text{1D}}|e^{2\pi i\hat{P}}|\Phi^{\text{1D}}\rangle|= e^{-\frac{2\pi^2 n^{\text{1D}} \lambda^2}{L_1}+O(\frac{1}{L_1^2})}.\label{Resta1D}
\end{equation}
where $n^{\text{1D}}=N/L_1$ is the particle density and $\lambda>0$ is the localization length.  Now, let us form a $d$-dimensional insulator by a $(d-1)$-dimensional array of the identical 1D chains with the lattice constant $a_i$ in $i$-th direction. Given Eq.~\eqref{Resta1D}, we have
\begin{eqnarray}
|\langle\Phi|e^{2\pi i\hat{P}}|\Phi\rangle|&=&|\langle\Phi^{\text{1D}}|e^{2\pi i\hat{P}}|\Phi^{\text{1D}}\rangle|^{\frac{L_2}{a_2}\cdots \frac{L_d}{a_d}}\notag\\
&=& e^{-2\pi^2n \lambda^2\frac{V}{L_1^2}+O(\frac{V}{L_1^3})},\,\,\, n\equiv\tfrac{n^{\text{1D}}}{a_2\cdots a_d}.\label{Souza}
\end{eqnarray}
Therefore, the magnitude $|\langle\Phi|e^{2\pi i\hat{P}}|\Phi\rangle|$ \emph{vanishes} when $\frac{V}{L_1^2}=\frac{L_2\cdots L_d}{L_1}\rightarrow+\infty$.  In fact, Eq.~\eqref{Souza} is consistent with the higher-dimensional generalization of the localization length proposed in Ref.~\onlinecite{Souza}.

\section{Polarization of band insulators}
\label{sec:band}
All the discussions so far apply regardless of the presence or absence of interactions or disorders, as long as the stated assumptions hold. 
Now let us consider the special case of band insulators, namely systems of noninteracting fermions in a periodic potential, with the Fermi level lying in a band gap. As we will see, the polarization in this case can be formulated in terms of Berry phases of the Bloch function in the momentum space.
As we will see, we will find two inequivalent Berry phases
$\mathcal{P}^{\text{Bloch}}$ and $\tilde{\mathcal{P}}^{\text{Bloch}}$
for a band insulator, which correspond to $\mathcal{P}$ and $\tilde{\mathcal{P}}$ for a many-body insulator as introduced above.
The difference between $\mathcal{P}^{\text{Bloch}}$ and $\tilde{\mathcal{P}}^{\text{Bloch}}$ in band insulators was examined earlier~\cite{Kudin2007,Rhim}. In particular, the relation between the surface charge and the bulk quantities $\mathcal{P}^{\text{Bloch}}$ and $\tilde{\mathcal{P}}^{\text{Bloch}}$ was discussed in Ref.~\cite{Rhim}.
In this paper, we provide a unified picture on polarization and Berry phase
for both many-body and band insulators, with a particular emphasis on the polarization current.

\subsection{One dimension, single occupied band}
As an example, let us take $\hat{H}_\theta$ in Eq.~\eqref{H1} with a periodic potential $V_{x+a}=V_x$ and set $\theta=0$.
The Hamiltonian can be block-diagonalized by the Fourier transformation
\begin{equation}
\hat{c}_{x}\equiv\tfrac{1}{\sqrt{L/a}}\sum_{n=1}^{L/a}\hat{c}_{k_n,r} e^{i k_n x},\quad k_n\equiv\tfrac{2\pi}{L}n.\label{Fourier1}
\end{equation}
Here, we decomposed $x$ as $x=R+r$ with $R=0,a,2a,\cdots,L-a$ and $0\leq r<a$, i.e., $R$ labels the unit cell and $r$ is the position within a cell.  The Hamiltonian then reduces to $\hat{H}_{\theta=0}=\sum_{n=1}^{L/a}\hat{h}_{k_n}$, where
\begin{equation}
\hat{h}_k\equiv\int_0^adr\,\hat{c}_{k,r}^\dagger h_{k,r}\hat{c}_{k,r},\,\,\,  h_{k,r}\equiv\tfrac{-(\partial_r+ik)^2}{2m}+V_r.\label{HTB}
\end{equation}
Observe the formal similarity between Eqs.~\eqref{H1} and \eqref{HTB} --- they are exactly mapped onto each other by $L\leftrightarrow a$ and $\tfrac{\theta}{L}\leftrightarrow k$. As a result, we can formulate the polarization of the band insulators in parallel with the general theory discussed before. 

For simplicity, let us consider the case the lowest band is completely filled and the other bands, separated by a band gap from the lowest band, are empty.
Let $u_k(r)$ be the lowest energy eigenstate of $h_{k,r}$ under the `periodic boundary condition' $u_k(r)=u_k(r+a)$.
We impose the normalization condition $\int_0^a dr | u_k(r)|^2 = 1$ for each $k$.
In analogy to Eq.~\eqref{manybodyP}, the polarization of the band insulator can be defined as~\cite{Zak,KSVPRB1993,VanderbiltKingSmith,RestaRMP,RestaVanderbilt}
\begin{equation}
\mathcal{P}^{\text{Bloch}}\equiv\int_{0}^{\frac{2\pi}{a}}\tfrac{dk}{2\pi}\int_0^adr\,iu_k(r)^*\partial_ku_k(r) .\label{PB1}
\end{equation}
The single-particle wavefunction in Fourier space, $u_k(r)$, can be interpreted as the periodic part of the Bloch function as we shall discuss shortly.  The seam-independence of $\mathcal{P}$ implies that $\mathcal{P}^{\text{Bloch}}$ does not depend on the choice of the unit cell~\cite{VanderbiltKingSmith,RestaVanderbilt}.
Note that $h_{k,r}$ and $u_{k}(r)$ are not periodic in $k$; they instead satisfy $h_{k+\frac{2\pi}{a},r}=e^{-2\pi i\frac{x}{a}}h_{k,r}e^{2\pi i\frac{r}{a}}$ and $u_{k+\frac{2\pi}{a}}(r)=e^{-2\pi i\frac{r}{a}}u_k(r)$~\cite{Zak,VanderbiltKingSmith,RestaVanderbilt,Rhim}.

\begin{table}
\caption{Comparison of Berry phases for 1D band insulators.  The atomic limit is the limit of the vanishing bopping with $\nu_i$ ($=0$ or $1$) localized electron at the site $x_i$.   (a)-(c) correspond to the model in Eq.~\eqref{fig1model} illustrated in Fig.~\ref{fig:1Dlattice} (a)-(c).
\label{summary12}}
\begin{tabular}{c|c|ccc}\hline\hline
			& Atomic Limit & (a) & (b) & (c)\\\hline
$\mathcal{P}^{\text{Bloch}}$ 		&  $\sum_{i}x_i\nu_i$	& 0.25	& 0.5		& 0.25 \\
$\tilde{\mathcal{P}}^{\text{Bloch}}$ 	&  0 					& 0		& 0		& $-0.5$ \\
$\bar{\mathcal{P}}_{0}^{\text{Bloch}}$&  $\sum_{i}x_i\nu_i$ 	& 0.25	& 0.5		& 0.75 \\\hline\hline
\end{tabular}
\end{table}

Just like there were two ways of describing the flux $\theta$, there are two equivalent conventions in the Fourier transformation. The alternative definition involves $e^{i k R}$ rather than $e^{i k x}$:
\begin{equation}
\hat{c}_{x}\equiv\tfrac{1}{\sqrt{L/a}}\sum_{n=1}^{L/a}\hat{\tilde{c}}_{k_n,r} e^{i k_n R}. \label{Fourier2}
\end{equation}
The two ways are simply related by $\hat{\tilde{c}}_{k,r}=e^{i k r}\hat{c}_{k,r}$.  In the latter choice, both $\tilde{h}_{k,r}=e^{ikr}h_{k,r}e^{-ikr}$ and $\tilde{u}_{k}(r)=e^{ikr}u_{k}(r)$ are manifestly periodic in $k$ with the period $2\pi/a$. For this reason, $\tilde{u}_{k}(r)$ is actually more standard in the context of topological insulators~\cite{TI2013}.   In turn, $\tilde{u}_{k}(r)$ satisfies the `twisted boundary condition' $\tilde{u}_{k}(r+a)=e^{ika}\tilde{u}_{k}(r)$.  The Berry phase with respect to $\tilde{u}_{k}(r)$
\begin{equation}
\tilde{\mathcal{P}}^{\text{Bloch}}\equiv\int_{0}^{\frac{2\pi}{a}}\tfrac{dk}{2\pi}\int_0^adr\,i\tilde{u}_k(r)^*\partial_k\tilde{u}_k(r)\label{PB2}
\end{equation}
measures the number of particles going through the unit-cell boundary.   The value of the fractional part of $\mathcal{P}^{\text{Bloch}}$ and $\tilde{\mathcal{P}}^{\text{Bloch}}$ do not agree, and we have $\mathcal{P}^{\text{Bloch}}=\tilde{\mathcal{P}}^{\text{Bloch}}+\bar{\mathcal{P}}_0^{\text{Bloch}}$~\cite{Rhim}, the analog of Eq.~\eqref{relP}, where
\begin{equation}
\bar{\mathcal{P}}_0^{\text{Bloch}}\equiv\int_{0}^{\frac{2\pi}{a}}\tfrac{dk}{2\pi}\int_0^adr\,|\tilde{u}_k(r)|^2r.\label{PB3}
\end{equation}

A possibly more familiar way~\cite{Zak,VanderbiltKingSmith,RestaVanderbilt} of introducing the same $u_{k}(r)$ and $\tilde{u}_{k}(r)$, without explicitly performing the Fourier transformation, is via the Bloch theorem.
It states that the single-particle wavefunction (the Bloch function) of $\hat{H}_{\theta=0}$ can be chosen in such a way that $\psi_{k}(x+a)=e^{ika}\psi_{k}(x)$ and $\psi_{k+\frac{2\pi}{a}}(x)=\psi_{k}(x)$.  Then it is customary to introduce the periodic part of the Bloch function via $u_{k}(r)=e^{-ikx}\psi_{k}(x)$. However, we could have defined $\tilde{u}_{k}(r)=e^{-ikR}\psi_{k}(x)$ as well. It is easy to see that $u_{k}(r)$ and $\tilde{u}_{k}(r)$ formulated this way agree with the ones above.

To see the properties of $\mathcal{P}^{\text{Bloch}}$ and $\tilde{\mathcal{P}}^{\text{Bloch}}$ more concretely, let us first discuss the atomic limit of tight-binding models.  In the limit of vanishing hopping, $\tilde{u}_k$ can always be chosen \emph{$k$-independent} so that $\tilde{\mathcal{P}}^{\text{Bloch}}=0$. This is expected since $\tilde{\mathcal{P}}^{\text{Bloch}}$ measures twist of $\tilde{u}_{k,i}$, and the atomic limit has the most trivial, constant Bloch function.  In contrast, $\mathcal{P}^{\text{Bloch}}$ may be nonzero even in this limit.  The $i$-th site at $x=R+r_i$ in each unit cell, if occupied, adds $r_i$ to $\mathcal{P}^{\text{Bloch}}$ via $u_{k,i}=e^{-i k r_i}$. More generally, if there is $\nu_i$ ($=0$ or $1$) localized electron at the site $x=R+r_i$, we get $\mathcal{P}^{\text{Bloch}}=\sum_{i}r_i\nu_i$ at the filling $\nu=\sum_{i}\nu_i$ in the atomic limit.   

As another simple exercise, let us examine a two-band model for the lattice in Fig.~\ref{fig:1Dlattice} (a):
\begin{equation}
\hat{H}=t_0\sum_{R/a=0}^{L/a-1}\hat{c}_{R}^{2\dagger}\hat{c}_{R}^1+\text{h.c.}\label{fig1model}
\end{equation}
The Hamiltonian contains only an intra-cell hopping $t_0>0$.  In this case, $\tilde{\mathcal{P}}^{\text{Bloch}}=0$ because the Hamiltonian in the momentum space does not have any $k$-dependence when the convention in Eq.~\eqref{Fourier2} is adopted.  However, if a different unit cell is chosen as in Fig.~\ref{fig:1Dlattice} (c), the same hopping becomes an inter-cell hopping and results in $\tilde{\mathcal{P}}^{\text{Bloch}}\neq0$. This clarifies the unit-cell dependence of $\tilde{\mathcal{P}}^{\text{Bloch}}$, translated from the seam-dependence of $\tilde{\mathcal{P}}$. In contrast, $\mathcal{P}^{\text{Bloch}}=0.25$ both for Fig.~\ref{fig:1Dlattice} (a) and (c) as summarized in Table~\ref{summary12}.

\subsection{General dimensions, multiple occupied bands}
For more general band insulators in $d$-dimensions with $\nu$-occupied bands, Eqs.~\eqref{PB1}, \eqref{PB2}, and \eqref{PB3} should be replaced by
\begin{align}
\mathcal{P}^{\text{Bloch}}_{\vec{k}_{\perp}} &=
\sum_{\alpha=1}^\nu
\int_0^{\frac{2\pi}{a_1}}\tfrac{d k_1}{2\pi}\int_{\text{u.c.}} d^d\vec{r}\,iu^{\alpha}_{\vec{k}}(\vec{r})^*\partial_{k_1}u^{\alpha}_{\vec{k}}(\vec{r}),
\\
\tilde{\mathcal{P}}^{\text{Bloch}}_{\vec{k}_{\perp}}&=\sum_{\alpha=1}^\nu\int_0^{\frac{2\pi}{a_1}}\tfrac{d k_1}{2\pi}\int_{\text{u.c.}} d^d\vec{r}\,i\tilde{u}^{\alpha}_{\vec{k}}(\vec{r})^*\partial_{k_1}\tilde{u}^{\alpha}_{\vec{k}}(\vec{r}),\\
\bar{\mathcal{P}}^{\text{Bloch}}_{0,\vec{k}_{\perp}}&=\sum_{\alpha=1}^\nu\int_0^{\frac{2\pi}{a_1}}\tfrac{d k_1}{2\pi}\int_{\text{u.c.}} d^d\vec{r}\,|u^{\alpha}_{\vec{k}}(\vec{r})|^2r_1.
\end{align}
where $u_{\vec{k}}^{\alpha}(\vec{r})=e^{-i\vec{k}\cdot\vec{r}}\tilde{u}_{\vec{k}}^{\alpha}(\vec{r})$ is the Bloch function of the $\alpha$-th occupied band,
$\vec{k}_{\perp}\equiv(k_2,\cdots,k_d)$ is the momentum perpendicular to $k_1$,  and $\text{u.c.}$ denotes the unit cell.

By computing the many-body Berry phases $\mathcal{P}$ and $\tilde{\mathcal{P}}$ in Eqs.~\eqref{manybodyP12} and~\eqref{tildeP}, which were defined for general interacting systems, for band insulators, 
we find
\begin{eqnarray}
\mathcal{P} &=&\tfrac{V}{L_1}\int\tfrac{d^{d-1}\vec{k}_{\perp}}{(2\pi)^{d-1}}\left[\tfrac{L_1-a_1}{2a_1}\nu+\mathcal{P}^{\text{Bloch}}_{\vec{k}_{\perp}}\right],
\label{PBgeneral}
\\
\tilde{\mathcal{P}}&=&\tfrac{V}{L_1}\int\tfrac{d^{d-1}\vec{k}_{\perp}}{(2\pi)^{d-1}}\tilde{\mathcal{P}}^{\text{Bloch}}_{\vec{k}_{\perp}},\\
\bar{\mathcal{P}}_0&=&\tfrac{V}{L_1}\int\tfrac{d^{d-1}\vec{k}_{\perp}}{(2\pi)^{d-1}}\left[\tfrac{L_1-a_1}{2a_1}\nu+\bar{\mathcal{P}}^{\text{Bloch}}_{0,\vec{k}_{\perp}}\right],
\end{eqnarray}
where the $\vec{k}_{\perp}$-integral is performed over $[0,\frac{2\pi}{a_2}]\times\cdots\times[0,\frac{2\pi}{a_d}]$.  See Appendix~\ref{app:BI} for the derivation.
These formulae may look ill-defined because of the factor $\tfrac{V}{L_1}$, but as we are only interested in the fractional part, divergence of the integer part does not really matter.

\section{Conclusions}
\label{sec:conclusion}

We have discussed several inequivalent Berry phases related to the polarization, and clarified their differences and mutual relations. Their values and behaviors can be quite different even for the same physical system, as it is evident in Fig.~\ref{fig:Delta} (b). Even the \emph{change} of the polarization over a generic time period [Eqs.~\eqref{dp1} and \eqref{dp2}], and thus the ``polarization current''  may depend on the definition. This difference can be attributed to the fact that they probe spatial regions differently, corresponding to the different gauges (distributions of the vector potential representing an Aharonov-Bohm flux).
Nevertheless, the total transported charge in a Thouless pumping process of one cyclic period is given by the same topological invariant (Chern number), independent of the Berry phase chosen for the calculation. 

The bulk polarization which is standard in the literature is $\mathcal{P}$ [Eq.~\eqref{manybodyP}] in the many-body context~\cite{OrtizMartin,AligiaOrtiz} or $\mathcal{P}^{\text{Bloch}}$ [Eq.~\eqref{PB1}]
for band insulators~\cite{Zak,KSVPRB1993,VanderbiltKingSmith,RestaRMP,RestaVanderbilt}, as $\mathcal{P}$ and $\mathcal{P}^{\text{Bloch}}$ take into account every point in space on the same footing.  However, we emphasize that other types of Berry phases are also well-defined. The polarization current derived from these Berry phases corresponds to the current measured locally.
The mutual relations between the inequivalent Berry phases are given, e.g., in Eq.~\eqref{relP}.
Once a certain definition is adopted, symmetries may quantize the possible values of the polarization. In such a case, the value of the polarization may be used to distinguish different phases, as it was done for example in Refs.~\onlinecite{Zak, NakamuraVoit, NakamuraTodo, HiranoKatsuraHatsugai1}.

\begin{acknowledgments}
H. W. thanks S. Murakami, K. Shiozaki, and M. Nakagawa for useful discussions on this topic.
M. O. thanks R. Kobayashi, Y.~O. Nakagawa, and Y. Fukusumi for a stimulating collaboration on a related project which motivated the present study, and G.-Y. Cho, M. Nakamura and G. Ortiz for useful discussions on related problems. In particular, we thank H. Katsura for fruitful discussions and comments on the initial draft, and J. W. Rhim for pointing out relevant references.
This work is supported in part by JSPS KAKENHI Grant Numbers JP17K17678 (H. W.) and JP16K05469 (M. O.).
\end{acknowledgments}

\bibliography{references}

\clearpage

\onecolumngrid

\appendix
 
\section{Current Induced by Adiabatic Time-Evolution of the Hamiltonian}
\label{app:thouless}
In this appendix we review the Thouless pump following Ref.~\onlinecite{NiuThouless}.  
Consider the adiabatic time-evolution of the density matrix $\hat{\rho}(t)=|\Psi(t)\rangle\langle\Psi(t)|$ over a long time-period $T$, obeying the time-dependent Schr\"odinger equation $i\hat{\rho}(t)=[\hat{H}(t),\hat{\rho}(t)]$.  At each time, the Hamiltonian $\hat{H}(t)$ has instantaneous eigenstates:
\begin{equation}
\hat{H}(t)|\Phi_n(t)\rangle=E_n(t)|\Phi_n(t)\rangle.
\end{equation}
Assuming that $|\Psi(0)\rangle$ is the ground state $|\Phi_0(0)\rangle$ at $t=0$ and that the time-evolution is adiabatic, let us introduce $\delta\hat{\rho}(t)\equiv\hat{\rho}(t)-|\Phi_0(t)\rangle\langle\Phi_0(t)|$. If the excitation gap remains finite at each time,  $E_n(t)-E_0(t)\geq\Delta>0$ ($n\neq0$), the adiabatic theorem says that $\|\delta\hat{\rho}(t)\|$ is of the order $(T\Delta)^{-1}$.

We want to find $\rho_{n0}(t)\equiv\langle\Phi_n(t)|\hat{\rho}(t)|\Phi_0(t)\rangle$ to the leading order by computing $\langle\Phi_n(t)|\partial_t\hat{\rho}(t)|\Phi_0(t)\rangle$ in two different ways. On the one hand, 
\begin{eqnarray}
\langle\Phi_n(t)|\partial_t\hat{\rho}(t)|\Phi_0(t)\rangle=-i\langle\Phi_n(t)|[\hat{H}(t),\hat{\rho}(t)]|\Phi_0(t)\rangle=-i[E_n(t)-E_0(t)]\rho_{n0}(t).
\end{eqnarray}
On the other hand, 
\begin{eqnarray}
\langle\Phi_n(t)|\partial_t\hat{\rho}(t)|\Phi_0(t)\rangle=(1-\delta_{n,0})\langle\Phi_n(t)|\partial_t|\Phi_0(t)\rangle+\langle\Phi_n(t)|[\partial_t\delta\hat{\rho}(t)]|\Phi_0(t)\rangle.
\end{eqnarray}
Thus we get the identity
\begin{equation}
\rho_{n0}(t)=\tfrac{i\langle\Phi_n(t)|\partial_t|\Phi_0(t)\rangle}{E_n(t)-E_0(t)}+\tfrac{i\langle\Phi_n(t)|[\partial_t\delta\hat{\rho}(t)]|\Phi_0(t)\rangle}{E_n(t)-E_0(t)}.\quad (n\neq0)
\end{equation}
The second term is of the order $(T\Delta)^{-2}$ and can be neglected.  

The expectation value of the local current density $\hat{j}_x\equiv\partial_{\theta}\hat{H}$ is thus given by
\begin{eqnarray}
j_x(t)&\equiv&\langle\Psi(t)|\hat{j}_x|\Psi(t)\rangle=\text{tr}[\hat{\rho}(t)\hat{j}_x]\notag\\
&\simeq&\langle\Phi_0(t)|\hat{j}_x|\Phi_0(t)\rangle+\sum_{n\neq0}\left[\langle\Phi_n(t)|\hat{j}_x|\Phi_0(t)\rangle \rho_{n0}^*+\langle\Phi_0(t)|\hat{j}_x|\Phi_n(t)\rangle\rho_{n0}
\right]\notag\\
&=&\langle\Phi_0(t)|\hat{j}_x|\Phi_0(t)\rangle-\sum_{n\neq0}i\left[\langle\Phi_n(t)|\hat{j}_x|\Phi_0(t)\rangle\tfrac{(\partial_t\langle\Phi_0(t)|)|\Phi_n(t)\rangle}{E_n(t)-E_0(t)}-\langle\Phi_0(t)|\hat{j}_x|\Phi_n(t)\rangle\tfrac{\langle\Phi_n(t)|\partial_t|\Phi_0(t)\rangle}{E_n(t)-E_0(t)}
\right]\notag\\
&=&\langle\Phi_0(t)|\partial_{\theta}\hat{H}|\Phi_0(t)\rangle-\sum_{n\neq0}i\left[(\partial_t\langle\Phi_0(t)|)|\Phi_n(t)\rangle\tfrac{\langle\Phi_n(t)|\partial_{\theta}\hat{H}|\Phi_0(t)\rangle}{E_n(t)-E_0(t)}-\langle\Phi_n(t)|\partial_t|\Phi_0(t)\rangle\tfrac{\langle\Phi_0(t)|\partial_{\theta}\hat{H}|\Phi_n(t)\rangle}{E_n(t)-E_0(t)}
\right]\notag\\
&=&\partial_{\theta}E_0(t)+\sum_{n\neq0}i\big[(\partial_t\langle\Phi_0(t)|)|\Phi_n(t)\rangle\langle\Phi_n(t)|\partial_{\theta}|\Phi_0(t)\rangle-(\partial_{\theta}\langle\Phi_0(t)|)|\Phi_n(t)\rangle\langle\Phi_n(t)|\partial_t|\Phi_0(t)\rangle
\big]\notag\\
&=&\partial_{\theta}E_0(t)+i\big[(\partial_t\langle\Phi_0(t)|)\partial_{\theta}|\Phi_0(t)\rangle-(\partial_{\theta}\langle\Phi_0(t)|)\partial_t|\Phi_0(t)\rangle\big].
\end{eqnarray}
In the derivation, we used 
\begin{eqnarray}
\langle\Phi_n(t)|\partial_{\theta}|\Phi_0(t)\rangle=-\tfrac{\langle\Phi_n(t)|\partial_{\theta}\hat{H}|\Phi_0(t)\rangle}{E_n(t)-E_0(t)}\quad (n\neq0)
\end{eqnarray}
and neglected $\rho_{00}-1$ and $\rho_{nm}$ ($n,m\neq0$) as they are of the order $(T\Delta)^{-2}$.

\section{Translation Symmetry of $\hat{H}_\theta$ and $\hat{\tilde{H}}_\theta$}
\label{app:twisted}
In this appendix we summarize the translation properties of $\hat{H}_\theta$ and $\hat{\tilde{H}}_\theta$.  Suppose that the Hamiltonian $\hat{H}_\theta$ is invariant under the translation symmetry $\hat{T}$.  Here, $\hat{T}$ is defined as the permutation operator:
\begin{equation}
\hat{T}\hat{c}_x\hat{T}^\dagger =
\begin{cases}
\hat{c}_{x+a} & (0\leq x<L-a),\\ 
\hat{c}_{x+a-L} & (L-a\leq x<L).
\end{cases}
\end{equation}
We have $\hat{T}^L=1$ by definition, which implies periodic boundary condition. 

The translation symmetry of $\hat{\tilde{H}}_{\theta}\equiv e^{i\theta\hat{P}}\hat{H}_\theta e^{-i\theta\hat{P}}$ reads
\begin{eqnarray}
\hat{\tilde{T}}_\theta&\equiv&e^{-i\theta\frac{a\hat{N}}{L}}e^{i\theta\hat{P}}\hat{T} e^{-i\theta\hat{P}}=e^{-i\theta\frac{a\hat{N}}{L}}e^{i\theta\hat{P}}e^{-i\theta(\hat{P}-\frac{a\hat{N}}{L}+\int_0^adx\hat{n}_x)}\hat{T}=e^{-i\theta\hat{n}_{R=0}}\hat{T},\
\end{eqnarray}
where $n_{R=0}\equiv\int_0^adx\hat{n}_x$.  The phase ambiguity of $\hat{\tilde{T}}_\theta$ is fixed in such a way that $\hat{\tilde{T}}_\theta$ becomes just the translation $\hat{T}$ except near the seam.  More explicitly,  we have
\begin{equation}
\hat{\tilde{T}}_\theta\hat{c}_x\hat{\tilde{T}}_\theta^\dagger=
\begin{cases}
\hat{c}_{x+a} & (0\leq x<L-a),\\ 
e^{-i\theta}\hat{c}_{x+a-L} & (L-a\leq x<L).
\end{cases}
\end{equation}
As a result, $\hat{\tilde{T}}_\theta$ satisfies $(\hat{\tilde{T}}_\theta)^L=e^{-i\theta \hat{N}}$, implying the twisted boundary condition.

\section{Current conservation}
\label{app:conservation}

Consider a Hamiltonian $\hat{H}_A$ under a \emph{classical} external field $A_x$.  For example, 
\begin{equation}
\hat{H}_A=\int_{0}^L dx\,\hat{c}_x^\dagger\left[-\tfrac{1}{2m}(\partial_x+iA_x)^2+V_x\right]\hat{c}_x.\label{appH1}
\end{equation}
Suppose that the Hamiltonian has the local U(1) symmetry. That is, under the gauge transformation $\hat{U}_\epsilon\hat{c}_x\hat{U}_\epsilon^\dagger=e^{-i\epsilon_x}\hat{c}_x$ with $\hat{U}_\epsilon\equiv e^{i\int dx\epsilon_x\hat{n}_x}$, $\hat{H}_A$ transforms as $\hat{U}_\epsilon\hat{H}_A\hat{U}_\epsilon^\dagger=\hat{H}_{\{A_x-\partial_x \epsilon_x\}}$.  We introduce the local current operator $\hat{j}_A(x)$ by $\hat{j}_A(x)\equiv\frac{\delta\hat{H}_A}{\delta A_x}$. 
Here, the functional derivative is defined by
\begin{eqnarray}
\hat{H}_{A+\delta A}=\hat{H}_A+\int_{0}^L dx\,\delta A_x\frac{\delta\hat{H}_A}{\delta A_x}+O((\delta A_x)^2).
\end{eqnarray}
For example, for the Hamiltonian in Eq.~\eqref{appH1}, $\hat{j}_A(x)=\tfrac{1}{2mi}\hat{c}_x^\dagger(\partial_x+iA_x)\hat{c}_x+\text{h.c}$.

Now take a small $\epsilon_x$ that vanishes at the boundary.  Define $\hat{H}\equiv\hat{H}_{A=0}$ and $\hat{j}(x)\equiv\hat{j}_{A=0}(x)$.  We get
\begin{eqnarray}
\hat{U}\hat{H}\hat{U}^\dagger&=&\hat{H}+i\int_{0}^L  dx\,\epsilon_x[\hat{n}_x,\hat{H}]+O(\epsilon_x^2),\\
\hat{U}\hat{H}\hat{U}^\dagger&=&\hat{H}_{-\partial_x \epsilon_x}=\hat{H}-\int_{0}^L dx\,\partial_x \epsilon_x\hat{j}(x)+O(\epsilon_x^2)=\hat{H}+\int_{0}^L dx\, \epsilon_x\partial_x\hat{j}(x)+O(\epsilon_x^2).
\end{eqnarray}
Since $\epsilon_x$ is arbitrary (except for the boundary condition), we get 
\begin{equation}
i[\hat{H},\hat{n}_x]+\partial_x\hat{j}(x)=0.
\end{equation}
In the Heisenberg picture $\partial_t\hat{n}_x=i[\hat{H},\hat{n}_x]$ and this is precisely the continuity equation. Furthermore, we have
\begin{eqnarray}
\partial_t\hat{P}(t)\equiv i[\hat{H},\hat{P}]=-\frac{1}{L}\int_{0}^L dx\,x\partial_x\hat{j}(x)=\frac{1}{L}\int_{0}^L dx\,\hat{j}(x)-\hat{j}(L)=\hat{\bar{j}}-\hat{j}(0).
\end{eqnarray}

\section{Delta-function potential}
\label{app:delta}

\subsection{Exact formulas for $\mathcal{N}_\theta$ and $\tilde{\mathcal{P}}$}
Here we provide two exact formulas on the model with a delta-function potential that were too long to be included in the main text. The normalization factor $\mathcal{N}_\theta$ reads
\begin{equation}
\mathcal{N}_\theta=\sqrt{\tfrac{q}{1-4qL e^{-2qL}-e^{-4qL}+4e^{-2qL}(qL\cosh qL-\sinh qL)\cos\theta}}.
\end{equation}
The Berry phase $\tilde{\mathcal{P}}$ is
\begin{equation}
\tilde{\mathcal{P}}=\tfrac{[\sqrt{2}qL\sinh(qL)-\sqrt{\cosh (2qL)-1-2(qL)^2}][\sinh(q(2\xi-L))+\tfrac{2\xi-L}{L}\sinh (qL)]}{2\sqrt{\cosh (2qL)-1-2(qL)^2}(qL\cosh(qL)-\sinh(qL))}-\tfrac{2\xi-L}{2L}.
\end{equation}

\subsection{Moving the seam}
\begin{figure}
\begin{center}
\includegraphics[width=0.40\columnwidth]{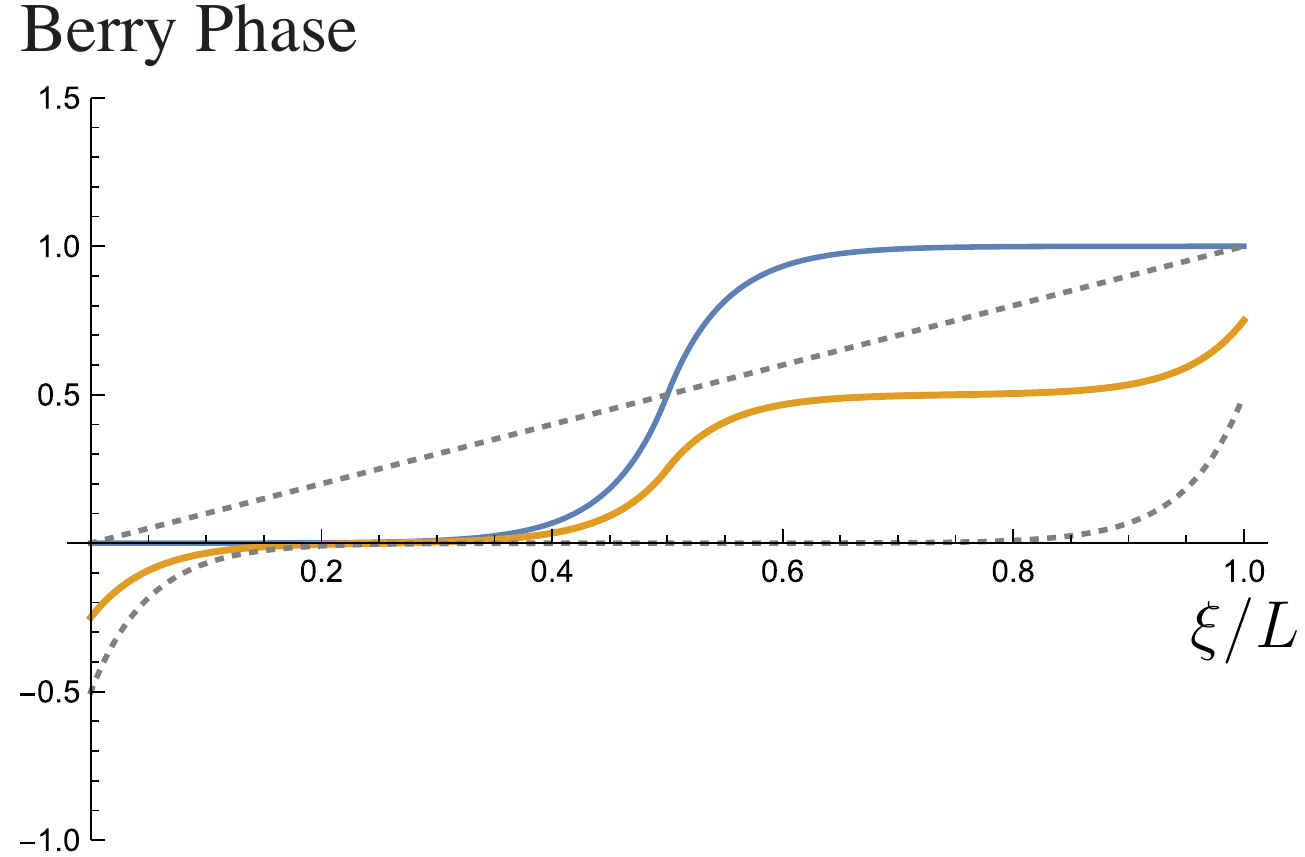}
\caption{\label{fig:Delta2} The behavior of the Berry phase for $\tilde{\Phi}_\theta(x)'$ (blue) and $\tilde{\Phi}_\theta(x)''$ (orange) as a function of $\xi$ for $qL=10$.}
\end{center}
\end{figure}

The position of the jump of $\mathcal{P}$ is not restricted to at $x=0$. If we move the seam by performing the gauge transformation $\tilde{\Phi}_\theta(x)'=e^{-i\theta\Theta(\frac{L}{2}<\theta<L)}\tilde{\Phi}_\theta(x)$ ($\Theta(x)=1$ when $x$ is true, and 0 otherwise), leading to $A_x(x)=\theta\delta(x-\frac{L}{2})$. The corresponding Berry phase has a jump at $x=\frac{L}{2}$. We can also split the jump into two positions by considering $\tilde{\Phi}_\theta(x)''=e^{-i\frac{1}{2}\theta\Theta(\frac{L}{2}<\theta<L)}\tilde{\Phi}_\theta(x)$, setting $A_x(x)=\frac{1}{2}\theta[\delta(x)+\delta(x-\frac{L}{2})]$. In this case, the Berry curvature measures the average of local current at $x=0$ and $x=\frac{L}{2}$. See Fig.~\ref{fig:Delta2}

\subsection{Thouless pump as the Aharonov-Bohm effect}
We can argue the pumped charge in a slightly different way. Let us introduce $\tilde{\tilde{\Phi}}_\theta(x-\xi)\equiv e^{i\theta\frac{\xi}{L}}\Phi_{\theta}(x)$, satisfying the periodicity in $\xi$ with the period $L$ but not in $\theta$:
\begin{equation}
\tilde{\tilde{\Phi}}_{\theta}(x)=\tilde{\tilde{\Phi}}_{\theta}(x-L), \quad \tilde{\tilde{\Phi}}_{0}(x)=e^{-i\frac{2\pi}{L}(x-\xi)}\tilde{\tilde{\Phi}}_{2\pi}(x).
\end{equation}
This wavefunction describes the bound state of $H_{\theta}(x-\xi)=-\tfrac{1}{2m}(\partial_x+i\tfrac{\theta}{L})^2-\tfrac{\lambda}{m}\delta(x-\xi)$. For this wavefunction, the Berry phase with respect to $\theta$ vanishes.  The Berry phase with respect to $\xi$ describes the Aharonov-Bohm phase~\cite{Berry}:
\begin{equation}
\mathcal{B}_\theta\equiv-\int_{0}^Ld\xi\int_{0}^Ldx\,i\tilde{\tilde{\Phi}}_{\theta}(x-\xi)^*\partial_\xi\tilde{\tilde{\Phi}}_{\theta}(x-\xi)=\theta+\tfrac{(qL)^2\sinh(qL)\sin\theta}{qL-\cosh(qL)\sinh(qL)+[\sinh(qL)-qL\cosh(qL)]\cos\theta}.
\end{equation}
The second term is a periodic function of $\theta$. The pumped charge is thus $\frac{\mathcal{B}_{\theta=2\pi}-\mathcal{B}_{\theta=0}}{2\pi}=+1$. 

\section{Resta's formula} 
\label{app:Resta}
\subsection{Estimate of $|\langle\Phi_{\theta}|\Phi_{\theta+d\theta}\rangle|$}
The first-order perturbation theory for $\hat{H}_{\theta+d\theta}=\hat{H}_{\theta}+d\theta\hat{\bar{j}}_\theta+O(d\theta)^2$ gives $E_{\theta+d\theta}=E_{\theta}+d\theta\langle\Phi_{\theta}|\hat{\bar{j}}_\theta|\Phi_{\theta}\rangle+O(d\theta)^2$ and
\begin{equation}
|\Phi_{\theta+d\theta}\rangle=\mathcal{N}_\theta\left[|\Phi_{\theta}\rangle-d\theta\hat{Q}_{\theta}\tfrac{1}{\hat{H}_\theta-E_\theta}\delta\hat{\bar{j}}_\theta|\Phi_{\theta}\rangle+O(d\theta)^2\right],\quad \delta\hat{\bar{j}}_\theta\equiv\hat{\bar{j}}_\theta-\langle\Phi_{\theta}|\hat{\bar{j}}_\theta|\Phi_{\theta}\rangle.
\end{equation}
Here,  $\hat{P}_{\theta}=|\Phi_{\theta}\rangle\langle\Phi_{\theta}|$ and $\hat{Q}_{\theta}=1-\hat{P}_{\theta}$ are the projection operators. The normalization factor $\mathcal{N}_\theta$ satisfies
\begin{eqnarray}
1\leq|\mathcal{N}_\theta|^{-2}&=&1+(d\theta)^2\langle\Phi_{\theta}|\delta\hat{\bar{j}}_\theta\tfrac{1}{\hat{H}_\theta-E_\theta}\hat{Q}_{\theta}\tfrac{1}{\hat{H}_\theta-E_\theta}\delta\hat{\bar{j}}_\theta|\Phi_{\theta}\rangle\leq1+\tfrac{(d\theta)^2}{\Delta_{\theta}^2}\langle\Phi_{\theta}|\delta\hat{\bar{j}}_\theta\hat{Q}_{\theta}\delta\hat{\bar{j}}_\theta|\Phi_{\theta}\rangle=1+(d\theta)^2\tfrac{\mathcal{C}_\theta V}{\Delta_{\theta}^2L_1^2},
\end{eqnarray}
where we introduced the normalized current fluctuation, which should converge to a $O(1)$ number in the limit of a large system size.
\begin{eqnarray}
\mathcal{C}_\theta\equiv \tfrac{L_1^2}{V}\langle\Phi_{\theta}|(\delta\hat{\bar{j}}_\theta)^2|\Phi_{\theta}\rangle=\tfrac{1}{V}\int_Vd^d\vec{x}d^d\vec{y}\langle\Phi_{\theta}|\delta\hat{j}_\theta(\vec{x})\delta\hat{j}_\theta(\vec{y})|\Phi_{\theta}\rangle.
\end{eqnarray}
Therefore, we get
\begin{equation}
1\geq|\mathcal{N}_\theta|^2=|\langle\Phi_{\theta}|\Phi_{\theta+d\theta}\rangle|^2\geq1-(d\theta)^2\tfrac{\mathcal{C}_\theta V}{\Delta_{\theta}^2L_1^2}\geq1-(d\theta)^2\tfrac{\mathcal{C}V}{\Delta^2L_1^2},
\end{equation}
where $\mathcal{C}$ is the maximum of $\mathcal{C}_\theta$ and $\Delta$ is the minimum of $\Delta_{\theta}$ as a function of $\theta$.

\subsection{Estimate of error in the evaluation of $\langle\Phi_{0}|\Phi_{2\pi}\rangle$}
If we approximate $\langle\Phi_{\theta-d\theta}|\Phi_{\theta+d\theta}\rangle$ by interpolating the mid-point, $\langle\Phi_{\theta-d\theta}|\Phi_{\theta}\rangle\langle\Phi_{\theta}|\Phi_{\theta+d\theta}\rangle$, the error $\varepsilon_{d\theta}\geq0$ is given by
\begin{eqnarray}
\varepsilon_{d\theta}=\langle\Phi_{\theta-d\theta}|\Phi_{\theta+d\theta}\rangle-\langle\Phi_{\theta-d\theta}|\Phi_{\theta}\rangle\langle\Phi_{\theta}|\Phi_{\theta+d\theta}\rangle=\langle\Phi_{\theta-d\theta}|(1-\hat{P}_{\theta})|\Phi_{\theta+d\theta}\rangle=\langle\Phi_{\theta-d\theta}|\hat{Q}_{\theta}|\Phi_{\theta+d\theta}\rangle.
\end{eqnarray}
Using the Schwartz inequality, we get
\begin{eqnarray}
\varepsilon_{d\theta}
&\leq&\sqrt{\langle\Phi_{\theta-d\theta}|\hat{Q}_{\theta}|\Phi_{\theta-d\theta}\rangle\langle\Phi_{\theta+d\theta}|\hat{Q}_{\theta}|\Phi_{\theta+d\theta}\rangle}\notag\\
&=&\sqrt{(1-\langle\Phi_{\theta-d\theta}|\hat{P}_{\theta}|\Phi_{\theta-d\theta}\rangle)(1-\langle\Phi_{\theta+d\theta}|\hat{P}_{\theta}|\Phi_{\theta+d\theta}\rangle)}\notag\\
&=&\sqrt{(1-|\mathcal{N}_{\theta-d\theta}|^2)(1-|\mathcal{N}_{\theta}|^2)}\leq1-(1-(d\theta)^2\tfrac{\mathcal{C}V}{\Delta^2L_1^2})=(d\theta)^2\tfrac{\mathcal{C}V}{\Delta^2L_1^2}
\end{eqnarray}
Therefore, by iteratively interpolating the mid-points, we get
\begin{eqnarray}
\langle\Phi_{0}|\Phi_{2\pi}\rangle&\simeq&\langle\Phi_{0}|\Phi_{\pi}\rangle\langle\Phi_{\pi}|\Phi_{2\pi}\rangle\simeq\langle\Phi_{0}|\Phi_{\frac{1}{2}\pi}\rangle\langle\Phi_{\frac{1}{2}\pi}|\Phi_{\pi}\rangle\langle\Phi_{\pi}|\Phi_{\frac{3}{2}\pi}\rangle\langle\Phi_{\frac{3}{2}\pi}|\Phi_{2\pi}\rangle\simeq\prod_{n=0}^{2^M-1}\langle \Phi_{\frac{2\pi n}{2^M}}|\Phi_{\frac{2\pi (n+1)}{2^M}}\rangle
\end{eqnarray}
after $M$ steps. The error of the $n$-th step is at most $2^{n-1}\varepsilon_{\frac{2\pi}{2^n}}$ and the total error can be bounded by
\begin{equation}
\sum_{n=1}^{\infty}2^{n-1}\varepsilon_{\frac{2\pi}{2^n}}=\tfrac{(2\pi)^2\mathcal{C}V}{2\Delta^2L_1^2}.
\end{equation}
Hence, the interpolating process can be verified in the limit $V/L_1^2\rightarrow0$.

\subsection{Band Insulator}
Here we confirm our understanding using the example of band insulators with $\nu$-occupied bands.   
To simplify expressions, let us introduce a shorthand notation for a $\nu$ by $\nu$ matrix $\langle u_{\vec{k}}|v_{\vec{k}}\rangle$:
\begin{equation}
\langle u_{\vec{k}}|v_{\vec{k}}\rangle_{\alpha',\alpha}=\int d^d\vec{r}\,u_{\vec{k}}^{\alpha'}(\vec{r})^*v_{\vec{k}}^\alpha(\vec{r}),
\end{equation}
where $\alpha=1,2\cdots,\nu$ is the band index of occupied bands.  Furthermore, we will write $\varphi=2\pi \tfrac{L_1-a_1}{2a_1}\tfrac{L_2}{a_2}\cdots\tfrac{L_d}{a_d}\nu$, $\vec{k}=(\frac{2\pi}{L_1}n_1,\frac{2\pi}{L_2}n_2,\cdots,\frac{2\pi}{L_d}n_d)$, and $\Delta\vec{k}=(\frac{2\pi}{L_1},0,\cdots,0)$.
With this notation, the ground state can be expressed as
\begin{eqnarray}
|\Phi_0\rangle&=&\prod_{\vec{k}}\prod_{\alpha=1}^\nu\gamma_{\vec{k},\alpha}^{\dagger}|0\rangle,\quad \hat{\gamma}_{\vec{k},\alpha}^\dagger=\int d^d\vec{r}\,u_{\vec{k}}^\alpha(\vec{r}) \hat{c}_{\vec{k},\vec{r}}^\dagger,\\
|\Phi_{2\pi}\rangle&=&e^{-i\varphi}\prod_{\vec{k}}\prod_{\alpha=1}^\nu\Big(\sum_{\alpha'=1}^\nu\langle u_{\vec{k}}|u_{\vec{k}+\Delta\vec{k}}\rangle_{\alpha',\alpha} \gamma_{\vec{k},\alpha'}^{\dagger}\Big)|0\rangle.
\end{eqnarray}
Writing the permutation of $\alpha$ as $\sigma(\alpha)$, we get
\begin{eqnarray}
\langle\Phi_{0}|\Phi_{2\pi}\rangle&=&e^{-i\varphi}\prod_{\vec{k}}\sum_{\sigma}\text{sign}\,\sigma\prod_{\alpha=1}^\nu\langle u_{\vec{k}}|u_{\vec{k}+\Delta\vec{k}}\rangle_{\sigma(\alpha),\alpha}=e^{-i\varphi}\prod_{\vec{k}}\text{det}\langle u_{\vec{k}}|u_{\vec{k}+\Delta\vec{k}}\rangle\notag\\
&=&e^{-i\varphi+\sum_{\vec{k}}\text{tr}\ln\langle u_{\vec{k}}|u_{\vec{k}+\Delta\vec{k}}\rangle}=e^{-i\varphi+\sum_{\vec{q}=\vec{k}+\frac{\Delta\vec{k}}{2}}\text{tr}\ln\langle u_{\vec{q}-\frac{\Delta\vec{k}}{2}}|u_{\vec{q}+\frac{\Delta\vec{k}}{2}}\rangle}\notag\\
&=&e^{-i\varphi+\sum_{\vec{q}}\text{tr}\left(\frac{\pi}{iL_1}\text{tr}C_{\vec{q},1}+\frac{1}{2}(\frac{\pi}{iL_1})^2\text{tr}C_{\vec{q},2}+\frac{1}{6}(\frac{\pi}{iL_1})^3\text{tr}C_{\vec{q},3}+\cdots\right)}\notag\\
&\simeq&e^{-i\varphi+V\int \frac{d^d\vec{q}}{(2\pi)^d}\,\text{tr}\left(-i\frac{\pi}{L_1}C_{\vec{q},1}-\frac{1}{2}(\frac{\pi}{L_1})^2C_{\vec{q},2}+\frac{1}{6}i(\frac{\pi}{L_1})^3C_{\vec{q},3}+\cdots\right)},\label{RestExp}
\end{eqnarray}
In the derivation, we defined $\text{tr}\ln\langle u_{\vec{q}-\frac{\Delta\vec{k}}{2}}|u_{\vec{q}+\frac{\Delta\vec{k}}{2}}\rangle=\tfrac{\pi}{iL_1}\text{tr}C_{\vec{q},1}+\tfrac{1}{2}(\tfrac{\pi}{iL_1})^2\text{tr}C_{\vec{q},2}+\tfrac{1}{6}(\tfrac{\pi}{iL_1})^3\text{tr}C_{\vec{q},3}+\cdots$, where
\begin{eqnarray}
C_{\vec{q},1}&\equiv& i(\langle u_{\vec{q}}|\partial_{q_x}u_{\vec{q}}\rangle-\langle \partial_{\vec{q}}u_{\vec{q}}|u_{\vec{q}}\rangle)=2i\langle u_{\vec{q}}|\partial_{q_x}u_{\vec{q}}\rangle,\\
C_{\vec{q},2}&\equiv& 4\langle \partial_{q_x}u_{\vec{q}}|\partial_{q_x}u_{\vec{q}}\rangle-C_{\vec{q},1}^2=4[\langle \partial_{q_x}u_{\vec{q}}|\partial_{q_x}u_{\vec{q}}\rangle-\langle \partial_{q_x}u_{\vec{q}}|u_{\vec{q}}\rangle\langle u_{\vec{q}}|\partial_{q_x}u_{\vec{q}}\rangle],\\
C_{\vec{q},3}&\equiv&-i(\langle u_{\vec{q}}|\partial_{q_x}^3u_{\vec{q}}\rangle-\langle \partial_{q_x}^3u_{\vec{q}}|u_{\vec{q}}\rangle-3\langle \partial_{q_x}u_{\vec{q}}|\partial_{q_x}^2u_{\vec{q}}\rangle+3\langle \partial_{q_x}^2u_{\vec{q}}|\partial_{q_x}u_{\vec{q}}\rangle)-3C_{\vec{q},1}C_{\vec{q},2}-C_{\vec{q},1}^3.
\end{eqnarray}
Note that $C_{\vec{q},\ell}$'s are all real.  The real part of $\ln\langle\Phi|e^{2\pi i\hat{P}}|\Phi\rangle$ is
\begin{eqnarray}
\text{Re}\ln\langle\Phi|e^{2\pi i\hat{P}}|\Phi\rangle&=&-\tfrac{\pi^2 V}{2L_1^2}\int\tfrac{d^d\vec{q}}{(2\pi)^d}\,\text{tr}C_{\vec{q},2}+O(\tfrac{V}{L_1^4})=-2\pi^2\tfrac{\Omega_{\text{I}}}{V_{\text{u.c.}}}\tfrac{V}{L_1^2}+O(\tfrac{V}{L_1^4}),\label{Resta1}
\end{eqnarray}
where $\Omega_{x}\equiv V_{\text{u.c.}}\int\tfrac{d^d\vec{q}}{(2\pi)^d}\,\text{tr}[\langle \partial_{q_x}u_{\vec{q}}|\partial_{q_x}u_{\vec{q}}\rangle-\langle \partial_{q_x}u_{\vec{q}}|u_{\vec{q}}\rangle\langle u_{\vec{q}}|\partial_{q_x}u_{\vec{q}}\rangle]$ is the spread of the Wannier function~\cite{MarzariVanderbilt} in $x$-direction and $V_{\text{u.c.}}=a_1\cdots a_d$ is the unit cell volume. Hence, since the particle density is related to the number of bands $\nu$ as $\nu=nV_{\text{u.c.}}$, we have $\nu\lambda^2=\Omega_{x}$.  The imaginary part of $\ln\langle\Phi|e^{2\pi i\hat{P}}|\Phi\rangle$ gives
\begin{eqnarray}
\mathcal{P}_{\text{R}}=\tfrac{1}{2\pi}\text{Im}\ln\langle\Phi|e^{2\pi i\hat{P}}|\Phi\rangle&=&\mathcal{P}-\tfrac{\pi^2 V}{12L_1^3}\int\tfrac{d^d\vec{q}}{(2\pi)^d}\,\text{tr}C_{\vec{q},3}+O(\tfrac{V}{L_1^5}).\label{Resta2}
\end{eqnarray}
Therefore, we have
\begin{equation}
\left|\langle\Phi|e^{2\pi i\hat{P}}|\Phi\rangle-e^{2\pi i\mathcal{P}}\right|=\left|e^{-2\pi^2n\lambda^2\frac{V}{L_1^2}+O(\frac{V}{L_1^3})}-1\right|=2\pi^2n\lambda^2\tfrac{V}{L_1^2}+O(\tfrac{V}{L_1^3},(\tfrac{V}{L_1^2})^2).
\end{equation}

In going to the last line of Eq.~\eqref{RestExp}, we approximated the discrete sum $\frac{2\pi}{L}\sum_{n=1}^{L/a}$ by an integral $\int_{0}^{\frac{2\pi}{a}}dk$. This process in general produces $O(L^{-1})$ corrections, but in fact the error is much smaller for smooth periodic functions~\cite{Trefethen}.  Let $f(k)$ be a smooth periodic function of $k$ with the period $\frac{2\pi}{a}$:
\begin{equation}
f(k)=\sum_{m=-\infty}^{\infty}f_me^{ika m},\quad f_m=\tfrac{a}{2\pi}\int_{0}^{\frac{2\pi}{a}}dke^{-ikam}f(k).
\end{equation}
If $\partial_k^Nf(k)=\sum_{m=-\infty}^{\infty}(iam)^Nf_me^{ika m}$ is finite for any $k$ and $N$, there exists $C$ and $\lambda$ such that $|f_m|\leq Ce^{-\lambda |m|}$. Then
\begin{eqnarray}
I_{L,\theta}\equiv\sum_{n=1}^{L/a}\tfrac{a}{L}f(k_n^\theta)=\sum_{n=1}^{L/a}\tfrac{a}{L}\sum_{m=-\infty}^{\infty}f_me^{i(\frac{2\pi}{L}n+\frac{\theta}{L})am}
=\sum_{\ell=-\infty}^{\infty}\sum_{m=-\infty}^{\infty}f_me^{i\frac{\theta}{L}am}\delta_{m,L\ell/a}=\sum_{\ell=-\infty}^{\infty}f_{L\ell/a}e^{i\theta\ell}.
\end{eqnarray}
Thus we get
\begin{eqnarray}
|I_{L,\theta}-f_0|\leq\sum_{\ell\neq0}|f_{L\ell/a}|\leq 2C\sum_{\ell=1}^\infty e^{-\lambda L\ell/a}=2C\tfrac{e^{-\lambda L/a}}{1-e^{-\lambda L/a}}.
\end{eqnarray}

\section{Many-Body Berry phases for Band Insulators}
\label{app:BI}
In this appendix, we compute many-body Berry phases $\mathcal{P}$, $\tilde{\mathcal{P}}$, $\bar{\mathcal{P}}_0$, and $\mathcal{P}_{\text{R}}$ for the Bloch Hamiltonian.  We consider a general situation of $\nu$-occupied bands in 1D.  The extension to higher dimensions is straightforward. 

\subsection{The ground state of $\hat{\tilde{H}}_\theta$}
The eigenvalues of $\hat{\tilde{T}}_\theta$ (see Appendix~\ref{app:twisted}) are $e^{-i k_n^\theta a}$ with $k_n^{\theta}\equiv k_n+\frac{\theta}{L}$.  The Fourier transformation is defined by
\begin{equation}
\hat{\tilde{c}}_{k_n^\theta,r}\equiv\tfrac{1}{\sqrt{L/a}}\sum_{R/a=0}^{L/a-1}\hat{c}_{x} e^{-i k_n^\theta R}=e^{i k_n^\theta r}\hat{c}_{k_n^\theta,r},\quad \hat{\tilde{T}}_\theta\hat{\tilde{c}}_{k_n^\theta,r} \hat{\tilde{T}}_\theta^\dagger=e^{i k_n^\theta a}\hat{\tilde{c}}_{k_n^\theta,r}.
\end{equation}

The single-particle Bloch state, generated by $\hat{\tilde{\gamma}}_{k_n^{\theta},\alpha}^\dagger=\int_0^adr\,\tilde{u}_{k_n^{\theta}}^\alpha(r) \hat{\tilde{c}}_{k_n^{\theta},r}^{\dagger}$, reads
\begin{equation}
\hat{\tilde{\gamma}}_{k_n^{\theta},\alpha}^\dagger|0\rangle=\int_0^Ldx\,\psi_{k_n^\theta}^\alpha(x)\hat{c}_x ^\dagger|0\rangle,\quad \psi_{k_n^\theta}^\alpha(x)=\tfrac{1}{\sqrt{L/a}}e^{ik_n^{\theta}R}\tilde{u}_{k_n^{\theta}}^\alpha(r).
\end{equation}
It satisfies $\hat{\tilde{T}}_\theta(\hat{\tilde{\gamma}}_{k_n^{\theta},\alpha}^\dagger|0\rangle)=e^{-ik_n^{\theta}}(\hat{\tilde{\gamma}}_{k_n^{\theta},\alpha}^\dagger|0\rangle)$ and $\psi_{k_n^\theta}^\alpha(x+L)=e^{i\theta}\psi_{k_n^\theta}^\alpha(x)$.  The insulating ground state of $\hat{\tilde{H}}_\theta$, fully occupying $\nu$-bands, can be written as
\begin{eqnarray}
|\tilde{\Phi}_{\theta}\rangle=e^{-i\frac{L-a}{2a}\nu\theta}\prod_{n=1}^{L/a}\prod_{\alpha=1}^\nu\hat{\tilde{\gamma}}_{k_n^{\theta},\alpha}^\dagger|0\rangle.\label{BIb}
\end{eqnarray}
When $\theta$ is increased from $0$ to $2\pi$, $k_n^{\theta}$ is shifted by $\frac{2\pi}{L}$ (i.e., $k_n^{2\pi}=k_{n+1}$) and the fermion operators should be rearranged to get back to the original ordering. The factor $(-1)^{(L/a-1)\nu}$ produced in this step is cancelled by the pre-factor in Eq.~\eqref{BIb} and $|\tilde{\Phi}_{\theta}\rangle$ satisfies the required periodicity in $\theta$,  $|\tilde{\Phi}_{\theta+2\pi}\rangle= |\tilde{\Phi}_{\theta}\rangle$.

\subsection{The ground state of $\hat{H}_\theta$}
The eigenvalues of $\hat{T}$ are $e^{-i k_n a}$.  The Fourier transformation is defined by
\begin{equation}
\hat{c}_{k_n,r}\equiv\tfrac{1}{\sqrt{L/a}}\sum_{R/a=0}^{L/a-1}\hat{c}_{x} e^{-i k_n x},\quad \hat{T}\hat{c}_{k_n,r} \hat{T}^\dagger=e^{i k_n a}\hat{c}_{k_n,r}.
\end{equation}
If we perform the unitary transformation, we get
\begin{equation}
e^{-i\theta\hat{P}}\hat{\tilde{c}}_{k_n^\theta,r}e^{i\theta\hat{P}}\equiv\tfrac{1}{\sqrt{L/a}}\sum_{R/a=0}^{L/a-1} \hat{c}_{x}e^{-i k_n^\theta R}e^{i \frac{R+r}{L}\theta}=e^{i k_n^\theta r}\hat{c}_{k_n,r}.
\end{equation}
The single-particle Bloch state reads
\begin{eqnarray}
&&\hat{\gamma}_{k_n^{\theta},\alpha}^\dagger\equiv e^{-i\theta\hat{P}}\hat{\tilde{\gamma}}_{k_n^{\theta},\alpha}^\dagger e^{i\theta\hat{P}}=\int_0^adr\,u_{k_n^{\theta}}^\alpha(r) \hat{c}_{k_n,r}^\dagger,\quad u_{k_n^{\theta}}^\alpha(r)=e^{-i k_n^\theta r}\tilde{u}_{k_n^{\theta}}^\alpha(r),\\
&&\hat{\gamma}_{k_n^{\theta},\alpha}^\dagger|0\rangle=\tfrac{1}{\sqrt{L/a}}\int_0^Ldx\,\psi_{k_n}^{\alpha'}(x)\hat{c}_x ^\dagger|0\rangle,\quad \psi_{k_n}^{\alpha'}(x)\equiv\psi_{k_n^\theta}^\alpha(x)e^{-i\frac{\theta x}{L}}=\tfrac{1}{\sqrt{L/a}}e^{ik_n^{\theta}x}u_{k_n^{\theta}}^\alpha(r)e^{-i\frac{\theta x}{L}}.
\end{eqnarray}
The insulating ground state of $\hat{H}_\theta$ is given by
\begin{eqnarray}
|\Phi_{\theta}\rangle=e^{-i\frac{L-a}{2a}\nu\theta}\prod_{n=1}^{L/a}\prod_{\alpha=1}^\nu\hat{\gamma}_{k_n^{\theta},\alpha}^\dagger|0\rangle.
\end{eqnarray}

\subsection{Berry Phases}
Given the ground states, we compute the many-body Berry phases. Let us start with $\mathcal{P}$:
\begin{eqnarray}
\mathcal{P}
&=&\tfrac{L-a}{2a}\nu+\sum_{n=1}^{L/a}\sum_{\alpha=1}^\nu\int_0^{2\pi}\tfrac{d\theta}{2\pi}\,i\langle0|\hat{\gamma}_{k_n^{\theta},\alpha}\partial_\theta\hat{\gamma}_{k_n^{\theta},\alpha}^\dagger|0\rangle=\tfrac{L-a}{2a}\nu+\sum_{n=1}^{L/a}\sum_{\alpha=1}^\nu\int_0^{2\pi}\tfrac{d\theta}{2\pi}\int_0^adr\,i u_{k_n^{\theta}}^\alpha(r)^* \partial_\theta u_{k_n^{\theta}}^\alpha(r)\notag\\
&=&\tfrac{L-a}{2a}\nu+\sum_{\alpha=1}^\nu\int_{0}^{\frac{2\pi}{a}}\tfrac{dk}{2\pi}\int_0^adr\,i u_{k}^\alpha(r)^* \partial_k u_{k}^\alpha(r)\equiv\tfrac{L-a}{2a}\nu+\mathcal{P}^{\text{Bloch}}.
\end{eqnarray}

Next, let us examine $\tilde{\mathcal{P}}$.
\begin{eqnarray}
\tilde{\mathcal{P}}
&=&\tfrac{L-a}{2a}\nu+\sum_{n=1}^{L/a}\sum_{\alpha=1}^\nu\int_0^{2\pi}\tfrac{d\theta}{2\pi}\,i\langle0|\hat{\tilde{\gamma}}_{k_n^{\theta},\alpha}\partial_\theta\hat{\tilde{\gamma}}_{k_n^{\theta},\alpha}^\dagger|0\rangle\notag\\
&=&\tfrac{L-a}{2a}\nu+\sum_{n=1}^{L/a}\sum_{\alpha=1}^\nu\int_0^{2\pi}\tfrac{d\theta}{2\pi}\int_0^adr\,i \tilde{u}_{k_n^{\theta}}^\alpha(r)^* \partial_\theta \tilde{u}_{k_n^{\theta}}^\alpha(r)\notag\\
&&\quad-\sum_{n=1}^{L/a}\sum_{\alpha=1}^\nu\int_0^{2\pi}\tfrac{d\theta}{2\pi}
\tfrac{a}{L}
\int_0^adr\int_0^adr'\,\tilde{u}_{k_n^{\theta}}^\alpha(r')^*\tilde{u}_{k_n^{\theta}}^\alpha(r) \sum_{R/a=0}^{L/a-1}\sum_{R'/a=0}^{L/a-1}\tfrac{R}{L}
e^{ik_n^\theta (R-R')}\langle0|\hat{c}_{R'+r'}\hat{c}_{R+r}^\dagger |0\rangle\notag\\
&=&\tfrac{L-a}{2a}\nu+\sum_{\alpha=1}^\nu\int_{0}^{\frac{2\pi}{a}}\tfrac{dk}{2\pi}\int_0^adr\,i \tilde{u}_{k}^\alpha(r)^* \partial_k \tilde{u}_{k}^\alpha(r)-\sum_{\alpha=1}^\nu\tfrac{L-a}{2}\int_0^{\frac{2\pi}{a}}\tfrac{dk}{2\pi}\int_0^adr\,|\tilde{u}_{k}^\alpha(r)|^2\notag\\
&=&\sum_{\alpha=1}^\nu\int_{0}^{\frac{2\pi}{a}}\tfrac{dk}{2\pi}\int_0^adr\,i \tilde{u}_{k}^\alpha(r)^* \partial_k \tilde{u}_{k}^\alpha(r)\equiv\tilde{\mathcal{P}}^{\text{Bloch}},
\end{eqnarray}
where we used $\langle0|\hat{c}_{R'+r'}\hat{c}_{R+r}^\dagger |0\rangle=\delta_{R,R'}\delta(r-r')$ and $\int_0^adr\,|\tilde{u}_{k}^\alpha(r)|^2=1$.  

On the other hand, $\bar{\mathcal{P}}_0$ is given by
\begin{eqnarray}
\bar{\mathcal{P}}_0&=&\int_0^{2\pi}\tfrac{d\theta}{2\pi}\int_0^{a}dr\sum_{R/a=0}^{L/a-1}\,\tfrac{r+R}{L}\langle\Phi_\theta|\hat{n}_{r+R}|\Phi_\theta\rangle=\int_0^{2\pi}\tfrac{d\theta}{2\pi}\int_0^{a}dr\sum_{R/a=0}^{L/a-1}\,\tfrac{r+R}{L}\langle\Phi_\theta|\hat{n}_{r}|\Phi_\theta\rangle\notag\\
&=&\int_0^{2\pi}\tfrac{d\theta}{2\pi}\int_0^{a}dr\,\tfrac{r}{a}\langle\Phi_\theta|\hat{n}_{r}|\Phi_\theta\rangle+\sum_{R/a=0}^{L/a-1}\,\tfrac{R}{L}\nu=\sum_{\alpha=1}^\nu\int_0^{\frac{2\pi}{a}}\tfrac{dk}{2\pi}\int_0^{a}dr\,|\tilde{u}_{k}^\alpha(r)|^2r+\tfrac{L-a}{2a}\nu\equiv\bar{\mathcal{P}}_0^{\text{Bloch}}+\tfrac{L-a}{2a}\nu.
\end{eqnarray}

These results are consistent with the following relation shown in the main text:
\begin{equation}
\mathcal{P}-\tilde{\mathcal{P}}-\bar{\mathcal{P}}_0=\mathcal{P}^{\text{Bloch}}-\tilde{\mathcal{P}}^{\text{Bloch}}-\bar{\mathcal{P}}_0^{\text{Bloch}}=0.
\end{equation}

\end{document}